\documentclass{article}
\usepackage{amsmath}
\usepackage{graphicx}
\usepackage{subcaption}
\usepackage{amsfonts} %conjuntos
\usepackage{amssymb}
\usepackage{authblk} %affil en el título
\usepackage{bm}
\usepackage{gensymb}
\usepackage{mathtools, nccmath}
\usepackage{float}
\usepackage{cancel}
\usepackage{hyperref}
\usepackage{notoccite}
\usepackage{physics}
\usepackage[acronym]{glossaries}
\usepackage{xcolor}
\usepackage{cancel}
\usepackage{url}
\usepackage[font={it}]{caption}
\usepackage{array,multirow,graphicx}
\usepackage{hhline}
\usepackage{booktabs}
\usepackage{indentfirst} 
\usepackage{cite}
\usepackage[a4paper,top=2cm,left=3cm,right=2cm,bottom=2.5cm]{geometry}
\title{\textbf{ The nuclear matter density functional under the nucleonic hypothesis}}
\author{Hoa Dinh Thi $^{1}$, Chiranjib Mondal $^{1}$ and Francesca Gulminelli $^{1,}$*}
\affil{$^{1}$ \textit{Laboratoire de Physique Corpusculaire, CNRS, ENSICAEN, UMR6534, Université de Caen Normandie, F-14000, Caen Cedex, France}}
\date{\today}
\begin{document}
	\maketitle
	\begin{abstract}
			A Bayesian analysis of the possible behaviors of the dense matter equation of state informed through recent 
		LIGO-Virgo as well as NICER measurements reveals that all the present observations are compatible with a fully 
		nucleonic hypothesis for the composition of dense matter, even in the core of the most massive pulsar PSR J0740+6620. 
		Under the hypothesis of a nucleonic composition, we extract the most general behavior of the energy per particle of 
		symmetric matter and density dependence of the symmetry energy, compatible with the astrophysical observations as well 
		as  our present knowledge of low energy nuclear physics from effective field theory predictions and experimental nuclear 
		mass data. These results can be used as a null hypothesis to be confronted with future constraints on dense matter to 
		search for possible exotic degrees of freedom.
	\end{abstract}
\section{Introduction}

The exceptional progress of multi-messenger astronomy on different astrophysical sources of dense matter has very recently  led to quantitative measurements of various properties of neutron stars (NS), such as the correlation between mass and radius (M-R) from X-ray timing with NICER \cite{nicer1,nicer2,nicer3, nicer4} and the tidal polarizability from gravitational wave (GW) LIGO/Virgo data \cite{ligo1,ligo2,ligo3,ligodec,virgodec}.  These observations, together with the plethora of upcoming data \cite{ligo4},  are expected to unveil in the next future exciting open questions such as the structure and degrees of freedom of baryonic matter in extreme  conditions, and in particular the presence of phase transitions and the existence of deconfined matter in the core of neutron stars \cite{Oertel2017}. 

This direct connection between astrophysical measurements and the microphysics of dense matter is due to the well-known fact that, under the  realm of general relativity, there is a one-to-one correspondence between any static observable and the dense matter equation of state (EoS) \cite{Hartle1967}.
However, the task is complicated by the fact that there is no ab-initio calculation of ultra-dense matter neither in the hadronic nor in the partonic sectors, and  therefore effective models are used. The information about the  composition of high density matter is blurred by the uncertainty on the effective energy functional, and similar equations of state can be obtained under different hypotheses on the underlying microphysics \cite{Horowitz2019,Burgio2020}.  

A  tension was reported between the GW observational data, that tend to favor stiffer EoS, and ab-initio nuclear physics calculations,  which point towards a slightly softer density dependence \cite{Guven2020}. Such a tension could in principle suggest the emergence of new degrees of freedom at high density. However, the statistical significance of the 
dispersion is not sufficient to lead to strong conclusions, and could even be reduced if the new measurement of the neutron-skin thickness of $^{208}Pb$ by the PREX-II collaboration \cite{prex} will confirm a higher value for the skin than previously estimated \cite{Piekarewicz2021,Essick2021}. In addition to that, the most recent M-R estimations from the two objects PSR J0740+6620 and PSR J0030+0451 do not report any significant reduction of the NS radius with increasing mass \cite{nicer3, nicer4}, in qualitative agreement with the expectations for purely hadronic models for the EoS \cite{metamodel_2}.

For these reasons, the hypothesis of a purely nucleonic composition of the NS cores cannot be ruled out. To identify the observables pointing towards more exotic constituents, it is important to quantitatively evaluate the space of parameters and observables  compatible with the nucleonic hypothesis.  
To this aim, meta-modelling techniques were proposed \cite{Steiner2013,metamodel_1,metamodel_2,Li2018,Lim2019,Carreau2019,Tsang2020}, which allow exploring the complete parameter space of hadronic equations of state, and predicting the astrophysical observables with uncertainties controlled by our present theoretical and experimental knowledge of nuclear physics. This approach can be viewed as a way to transform experimental and observational constraints into nuclear physics empirical parameters to guide the elaboration of phenomenological and microscopic nuclear models, and it
can also be used as a null hypothesis to search for exotic degrees of freedom.

In this paper, we address this timely issue by performing a Bayesian statistical analysis of the semi-agnostic 
meta-modelling technique  of Refs. \cite{metamodel_1,Carreau2019}, including both nuclear physics and astrophysical 
constraints. With respect to previous works by different groups  \cite{Steiner2013,metamodel_1,metamodel_2,Li2018,
	Lim2019,Carreau2019,Tsang2020}, we include the most recent NICER results \cite{nicer4} which  give  
constraints in the density region where many-body 
perturbation theory (MBPT) cannot be applied, and use a fully unified EoS approach \cite{CarreauEPJA} allowing to 
include in the posterior probabilities the constraints coming from nuclear mass measurements \cite{ame2016}. 
 We have not included in the considered constraints the recent skin measurement by PREX II \cite{prex} 
	because our model is not presently able to calculate nuclear radii. An extension in this direction would be 
	of foremost interest, and it is left for future work.

The paper is organized as follows. In section 2, we summarize the basic ideas of nucleonic metamodelling 
developed in Ref. \cite{metamodel_1,metamodel_2}. We explain the different filters from low energy nuclear physics 
and astrophysical observations used for the Bayesian analysis in Section 3. The results obtained in the present work 
are described in Section 4. We make our concluding remarks in Section 5. 

\section{Meta-modelling of the EoS} \label{sec:model}

Within the assumption that the core of neutron stars is composed of neutrons, protons, electrons, and muons in weak equilibrium, a prior distribution of viable unified EoS model is generated by Monte-Carlo sampling of a large parameter set of 10 independent, uniformly distributed empirical parameters  corresponding to the successive density derivatives at saturation  up to order 4 of the uniform matter binding energy in the isoscalar and isovector channels. These parameters characterize the density dependence of the energy in symmetric matter and of the symmetry energy, and their prior distribution is consistent with the present empirical knowledge for a large set of nuclear data \cite{metamodel_1}. They
are complemented by 5 additional surface and curvature parameters \cite{Hoa2021} that are optimized, for each set of uniform matter parameters, to the experimental Atomic Mass Evaluation 2016 (AME2016) nuclear mass table  \cite{ame2016}. 
 The expression of the surface and curvature energy we employ \cite{Ravenhall1983}  was optimized on  
	Thomas-Fermi calculations at extreme isospin asymmetries, also subsequently employed in different works on 
	neutron star crust and supernova modelling within the compressible liquid drop approximation 
	\cite{Lattimer1991, Lorenzet93, Newton2013, CarreauEPJA, Balliet21, Hoa2021}.
{Two additional  parameters rule the density dependence of the effective mass and the effective mass splitting, and an extra parameter enforces the correct behavior at zero density, see Ref. \cite{metamodel_1} for details.} The use of the same functional to describe the inhomogeneous crust \cite{CarreauEPJA,Hoa2021} guarantees a consistent estimation of the crust-core transition and is known to be important for a correct estimation of the NS radius \cite{Fortin2016}.

\section{Bayesian analysis} \label{sec:Bayes}

The posterior distributions of the set $\bf X$ of EoS parameters  are conditioned by likelihood models of the different observations and constraints $\mathbf c $ according to the standard definition:
%% If the documentclass option "submit" is chosen, please insert a blank line before and after any math environment (equation and eqnarray environments). This ensures correct linenumbering. The blank line should be removed when the documentclass option is changed to "accept" because the text following an equation should not be a new paragraph.

\begin{equation}
	P({\mathbf X}|{\mathbf c})=\mathcal N   P({\mathbf X}) \prod_k P(c_k|{\mathbf X}),
\end{equation}
where $ P({\mathbf X})$ is the prior, and $\mathcal N$ is a normalization factor. The different constraints $c_k$ used in the present study are the following: 
(a) nuclear mass measurements in the AME2016 mass table \cite{ame2016}; (b)  the bands of allowed region in symmetric and pure neutron matter produced by many-body perturbation theory (MBPT) calculations from Ref. \cite{Drischler2016} based on two- and three-nucleon chiral effective field-theory (EFT) interactions at  next-to-next-to-next-to leading order (N3LO),  that are interpreted as a  $90\%$ confidence interval;  (c) mass measurement from radio-timing observations of pulsar PSR J0348+0432 \cite{Antoniadis2013}, $M_{J03} = 2.01 \pm 0.04 M_{\odot}$, where $M_{\odot}$ is the solar mass; (d) constraints on the tidal deformability of the binary NS system associated to the gravitational wave event GW170817, detected by the LIGO/Virgo Collaboration (LVC) \cite{ligo3}; (e) X-ray pulse-profile measurements of PSR J0030+0451’s  mass, $M_{J00}=1.44^{+0.15}_{-0.14} M_{\odot}$,   and radius, $R_{J00}=13.02^{+1.24}_{-1.06}$ km from Ref. \cite{nicer2}; (f) the radius measurement with NICER and XMM-Newton data  \cite{nicer4} of  the  PSR J0740+6620 pulsar  of mass $M_{J07}=2.08 \pm 0.07 M_{\odot}$ \cite{Fonseca2021}, $R_{J07}=13.7^{+2.6}_{-1.5}$ km  \cite{nicer4}. 

Posterior distributions of different observables $Y$ are  calculated by marginalizing over the EoS parameters  as:
%% If the documentclass option "submit" is chosen, please insert a blank line before and after any math environment (equation and eqnarray environments). This ensures correct linenumbering. The blank line should be removed when the documentclass option is changed to "accept" because the text following an equation should not be a new paragraph.

\begin{equation}
	P(Y|{\mathbf c})=\prod_{k=1}^N\int_{X_k^{min}}^{X_k^{max}}dX_k \, P({\mathbf X}|{\mathbf c}) \delta\left (Y-Y({\mathbf X})\right ),
\end{equation}
where $N=13$ is the number of parameters in the metamodel.  $Y({\mathbf X})$ is the value of   any observable $Y$ obtained with the ${\mathbf X}$ parameter set,  with $X_k^{\rm min(max)}$ being the minimum (maximum) value in the prior distribution taken as in Ref. \cite{CarreauEPJA}. In order to see the impact of different constraints on the nuclear physics informed prior, we consider  four distributions, each containing around $\sim$ 18000 models. They are labeled as follows:

\begin{enumerate}
	\item {\it Prior}: models in this set are required to result in meaningful solutions for the crust, that is, the minimization of the canonical thermodynamic potential at a given baryon density leads to positive gas and cluster densities. In addition, the fit of the surface and curvature parameters $\{\sigma_0,b_s, \sigma_{0c}, \beta\}$ to the nuclear masses in the AME2016 table must be convergent. These criteria are characterized by the pass-band filter $\omega_0$. Given that the mentioned conditions are satisfied, i.e., $\omega_0=1$, the probability of each model, associated to a parameter set ${\bf X}$, is then quantified by the goodness of the optimal fit,
	%% If the documentclass option "submit" is chosen, please insert a blank line before and after any math environment (equation and eqnarray environments). This ensures correct linenumbering. The blank line should be removed when the documentclass option is changed to "accept" because the text following an equation should not be a new paragraph.
	
	\begin{equation}
		P_{1}(\mathbf X) \propto \omega_0 e^{-\chi^2({\bf X})/2} P({\mathbf X}),
		\label{eq:P_prior}
	\end{equation}
	
	in which the original prior $P({\mathbf X})$ contains uniformly distributed EoS parameters, and the cost function $\chi^2({\bf X})$ has the following form:
	%% If the documentclass option "submit" is chosen, please insert a blank line before and after any math environment (equation and eqnarray environments). This ensures correct linenumbering. The blank line should be removed when the documentclass option is changed to "accept" because the text following an equation should not be a new paragraph.
	
	\begin{equation}
		\chi^2({\bf X})=\frac{1}{N_{dof}}\sum_n \frac{\left ( M_{cl}^{(n)}({\bf X})-M_{AME}^{(n)}\right )^2}{\sigma_{n}^2}. 
		\label{eq:chi2}
	\end{equation}
	
	The sum in Eq. \ref{eq:chi2} runs over all the nuclei in the AME2016 \cite{ame2016} mass table; $M_{AME}$ and  $M_{cl}({\bf X})$  being the experimental  and theoretical nuclear masses, respectively, in which the latter is calculated within a compressible liquid drop model (CLDM) approximation using the best-fit surface and curvature parameters for each EOS model;  $\sigma_{n}$ represents the systematic theoretical error; and $N_{dof}  (=n-4)$ is the number of degrees of freedom.  The distributions obtained with this prior represent the most general predictions, within a purely nucleonic composition hypothesis, that are compatible with low energy nuclear physics experiments.
	
	\item {\it LD}: in this sample the models are selected  by the strict filter from the chiral EFT calculation, where  the energy per nucleon of symmetric nuclear matter (SNM) and pure neutron matter (PNM) of the model are compared with the corresponding energy bands of Ref. \cite{Drischler2016}, enlarged by 5\%. This constraint is applied in the low-density region, from 0.02 fm$^{-3}$ to 0.2 fm$^{-3}$. The posterior probability can be written as:
	%% If the documentclass option "submit" is chosen, please insert a blank line before and after any math environment (equation and eqnarray environments). This ensures correct linenumbering. The blank line should be removed when the documentclass option is changed to "accept" because the text following an equation should not be a new paragraph.
	
	\begin{equation}
		P_{2}(\mathbf X) \propto \omega_{LD}(\mathbf X) P_{1}(\mathbf X),
		\label{eq:P_LD}
	\end{equation}
	
	in which $\omega_{LD}(\mathbf X)=1$ if the model $\bf X$ is consistent with the EFT  bands, and $\omega_{LD}(\mathbf X)= 0$ otherwise.  Implementing this low density (LD) filter amounts to including in the nucleonic hypothesis the information from ab-initio nuclear theory.
	
	\item {\it HD + LVC}: the posterior probability of this distribution is written as:
	%% If the documentclass option "submit" is chosen, please insert a blank line before and after any math environment (equation and eqnarray environments). This ensures correct linenumbering. The blank line should be removed when the documentclass option is changed to "accept" because the text following an equation should not be a new paragraph.
	
	\begin{equation}
		P_{3}({\bf X}) \propto \omega_{HD}P(J03|{\mathbf X})P(LVC|{\mathbf X})P_{1}(\mathbf X).
		\label{P_HDLVC}
	\end{equation}
	
	Here, $\omega_{HD}$ is also a pass-band type filter  similar to $\omega_{LD}$ in Eq. \ref{eq:P_LD}. It only accepts models satisfying all the following conditions: causality, thermodynamic stability, and non-negative symmetry energy at all densities. The second term in Eq. \ref{P_HDLVC}, $P(J03|{\mathbf X})$, is the likelihood probability from the mass measurement of  PSR J0348+0432 \cite{Antoniadis2013}, which is $M_{J03} = 2.01 \pm 0.04  M_{\odot}$. This likelihood is defined as the cumulative Gaussian distribution function  with a mean value of 2.01 and a standard deviation of 0.04:
	%% If the documentclass option "submit" is chosen, please insert a blank line before and after any math environment (equation and eqnarray environments). This ensures correct linenumbering. The blank line should be removed when the documentclass option is changed to "accept" because the text following an equation should not be a new paragraph.
	
	\begin{equation}
		P(J03|{\mathbf X}) =\frac{1}{0.04\sqrt{2\pi}} \int_{0}^{M_{max}({\bf X})/M_{\odot}}e^{-\frac{(x-2.01)^2}{2\times 0.04^2}}dx,
		\label{Eq:mmax}
	\end{equation}
	
	where $M_{max}({\bf X})$ is the maximum NS mass at equilibrium,  determined from the solution of the Tolmann-Oppenheimer-Volkoff (TOV) equations \cite{ Oppenheimer1939, Tolman1939}.
	
	We expect these different conditions not to be selective on the low order EOS parameters,
	but to constitute stringent constraints on the high density (HD) behavior of the EOS that is essentially governed, within the nucleonic hypothesis, by the third and fourth order effective parameters $Q_{sat}$, $Z_{sat}$, $Q_{sym}$ and $Z_{sym}$  \cite{metamodel_1}. 
	
	The constraint from the  GW170817 event, measured by the LVC,  evaluates the weight of a model based on its prediction for the tidal deformability $\tilde{\Lambda}$. The likelihood is written as :
	%% If the documentclass option "submit" is chosen, please insert a blank line before and after any math environment (equation and eqnarray environments). This ensures correct linenumbering. The blank line should be removed when the documentclass option is changed to "accept" because the text following an equation should not be a new paragraph.
	
	\begin{equation}
		P(LVC|{\mathbf X})= \sum_{i} P_{LVC}(\tilde{\Lambda}(q^{(i)}), q^{(i)}),
		\label{eq:LVC}
	\end{equation}
	
	in which $q$ is the ratio of the lighter component mass $m_2$ to the heavier component mass $m_1$, $q = m_2/m_1 \leq 1$, and $P_{LVC}(\tilde{\Lambda}(q), q)$ is the joint posterior distribution of $\tilde{\Lambda}$ and $q$ taken from Refs. \cite{ligo3, dataligo}. In  Refs. \cite{ligo3, dataligo}, the authors performed a Bayesian inference with four different waveform models.  The distribution for $\tilde{\Lambda}$ and $q$  which we are using for this work is the one obtained with the PhenomPNRT waveform, which is mentioned as their ``reference model". The tidal deformability $\tilde{\Lambda}$  is expressed in the form  of the component masses, $m_1$ and $m_2$, and the two corresponding dimensionless tidal deformabilities, $\Lambda_1$ and $\Lambda_2$, as :
	%% If the documentclass option "submit" is chosen, please insert a blank line before and after any math environment (equation and eqnarray environments). This ensures correct linenumbering. The blank line should be removed when the documentclass option is changed to "accept" because the text following an equation should not be a new paragraph.
	
	\begin{equation}
		\tilde{\Lambda}=\frac{16}{13}\frac{(m_1+12m_2)m_1^4\Lambda_1+(m_2+12m_1)m_2^4\Lambda_2}{(m_1+m_2)^5}.
	\end{equation}
	
	The dimensionless tidal deformability $\Lambda$ is related to the mass $M$ through the expression:
	%% If the documentclass option "submit" is chosen, please insert a blank line before and after any math environment (equation and eqnarray environments). This ensures correct linenumbering. The blank line should be removed when the documentclass option is changed to "accept" because the text following an equation should not be a new paragraph.
	
	\begin{equation}
		\Lambda = \frac{2}{3}k_2\left[\frac{c^2}{G}\frac{R(M)}{M}\right]^5,
		\label{eq:lambda}
	\end{equation}
	
	where $c$, $G$, $R(M)$, and $k_2$   are the speed of light, gravitational constant, NS radius at mass $M$, and  Love number, respectively \cite{Hinderer2008, Binnington2009, Damour2009}.
	In  our analysis, $q$ is chosen to be in the one-sided $90\%$ confidence interval obtained in Ref. \cite{ligo3}, $q \in \left[0.73, 1.00\right]$.  In Ref. \cite{ligo3}, it  was shown that  the chirp mass  $\mathcal{M}_c$ of the binary NS system associated to the GW170817 event was  accurately determined,  $\mathcal{M}_c=1.186\pm0.001 M_{\odot}$ at the median value with $90\%$ confidence limits. The chirp mass $\mathcal{M}_c$ can be expressed as a function of $m_1$ and $q$ as:
	%% If the documentclass option "submit" is chosen, please insert a blank line before and after any math environment (equation and eqnarray environments). This ensures correct linenumbering. The blank line should be removed when the documentclass option is changed to "accept" because the text following an equation should not be a new paragraph.
	
	\begin{equation}
		\mathcal{M}_c = \frac{(m_1m_2)^{3/5}}{(m_1+m_2)^{1/5}} =\frac{q^{3/5}m_1}{(1+q)^{1/5}}.
		\label{chirpmass}
	\end{equation}
	
	Since the uncertainty in the chirp mass  $\mathcal{M}_c$ is negligible, for each value of the mass ratio $q$, we calculate $m_1$ directly from the median value of $\mathcal{M}$ through Eq. (\ref{chirpmass}).
	%% If the documentclass option "submit" is chosen, please insert a blank line before and after any math environment (equation and eqnarray environments). This ensures correct linenumbering. The blank line should be removed when the documentclass option is changed to "accept" because the text following an equation should not be a new paragraph.
	
	%\begin{equation}
	%	\xout{	m_1 =} \frac{\xout{(1+q)^{1/5}}}{\xout{q^{3/5}}}\xout{\times %1.186 M_{\odot}.}
	%\end{equation}
	
	\item {\it All}: Including  the three constraints mentioned above together with the likelihood from the joint mass-radius distributions of the two NICER measurements from Refs. \cite{nicer2,nicer4}, the posterior probability for the final distribution is written as:
	%% If the documentclass option "submit" is chosen, please insert a blank line before and after any math environment (equation and eqnarray environments). This ensures correct linenumbering. The blank line should be removed when the documentclass option is changed to "accept" because the text following an equation should not be a new paragraph.
	
	\begin{equation}
		P_{4}(\mathbf X)   \propto 	\omega_{LD}\omega_{HD}P(J03|{\mathbf X})P(LVC|{\mathbf X}) P(NICER|{\mathbf X})P_{1}(\mathbf X).
	\end{equation}
	
	The NICER likelihood probability is given by:
	%% If the documentclass option "submit" is chosen, please insert a blank line before and after any math environment (equation and eqnarray environments). This ensures correct linenumbering. The blank line should be removed when the documentclass option is changed to "accept" because the text following an equation should not be a new paragraph.
	
	\begin{equation}
		P(NICER|{\mathbf X}) = \sum_{i}p_{NICER1}(M_1^{(i)}, R(M_1^{(i)})) \sum_{j}p_{NICER2}(M_2^{(j)}, R(M_2^{(j)})),
	\end{equation}
	
	where $p_{NICER1}(M,R)$ is the two-dimensional probability distribution of mass and radius for the pulsar PSR J0030+0451 obtained using the waveform model with three uniform oval spots by Miller \textit{et al}. in \cite{nicer2}; and $p_{NICER2}(M,R)$ is the probability distribution for  PSR J0740+6620  using NICER and XMM-Newton data by Miller \textit{et al}.   \cite{nicer4}. The intervals of $M_1$ and $M_2$ are chosen sufficiently large so that they cover most of the associated joint mass-radius distributions,  $M_1 \in \left[1.21,1.70\right] M_{\odot}$ and  $M_2 \in \left[1.90,2.25\right] M_{\odot}$.   
\end{enumerate}

 To insure that the differences in the posterior distributions are induced by the impact of the different 
	constraints,  care is taken to have a comparable statistics from the four distributions, for each plot shown in 
	this paper. Moreover, for all shown observables we have checked that an increase in statistics  does not affect 
	the results, within the precision chosen for the numerical values  given in this paper.

\section{Results and Discussions}
%All figures and tables should be cited in the main text as Figure~\ref{fig1}, Table~\ref{tab1}, etc.
\subsection{Empirical parameters}

\begin{figure}[!htpb]
	\centering
	\includegraphics[scale=0.27]{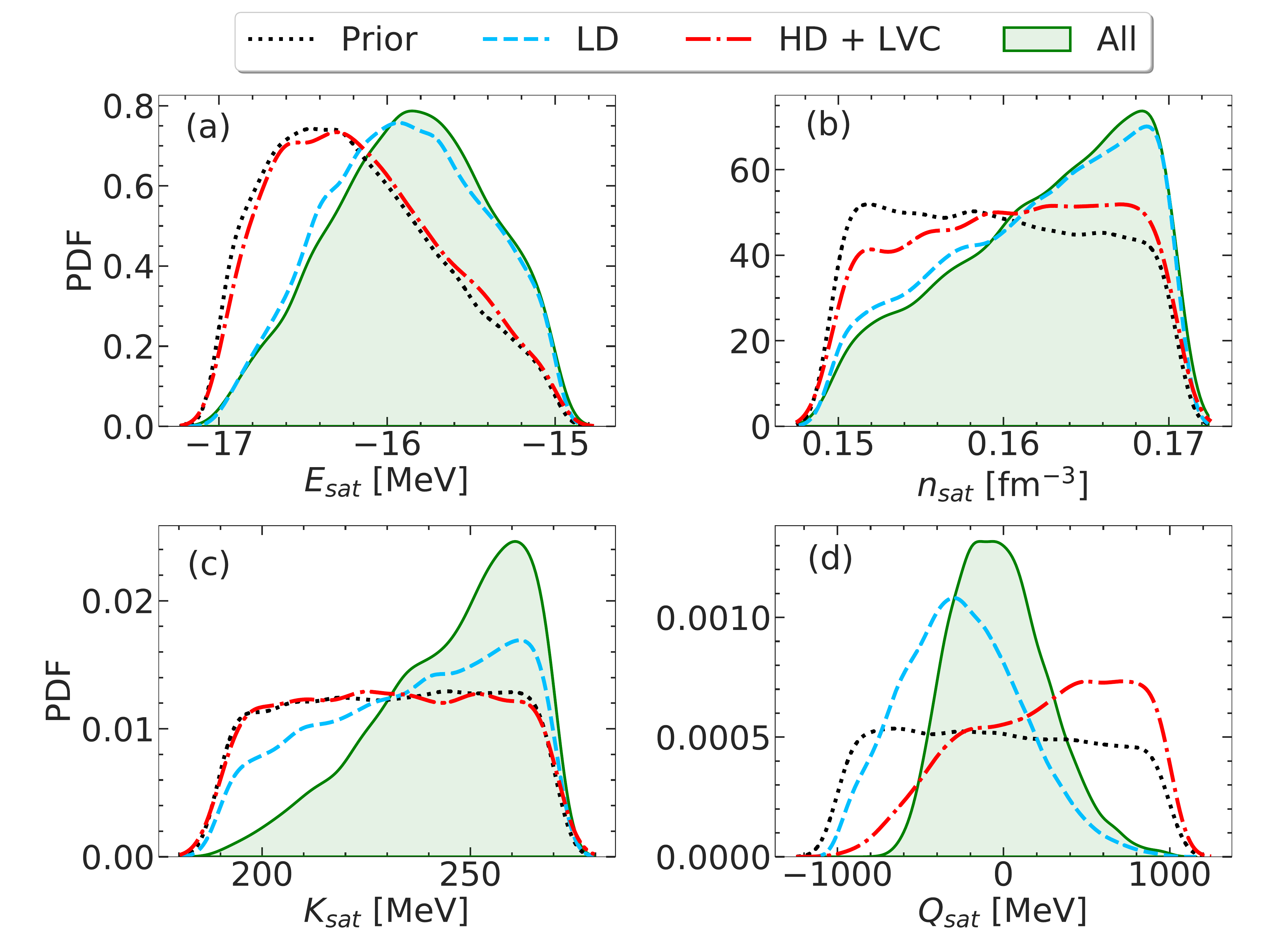}
	\caption{Probability density distributions of isoscalar empirical parameters for the prior distribution informed by experimental nuclear masses (black dotted line) and for posteriors of models passing through the low-density (chiral EFT) constraint (blue dashed line), high-density  constraints (causality, stability, $e_{sym} \geq 0$, maximum NS mass, and tidal deformability) (red dash-dotted line), and all constraints combined (green shaded region). See texts for details.  \label{isoscalar}}
\end{figure}

\begin{figure}[!htpb]
	\centering
	\captionsetup{justification=centering}
	\includegraphics[scale=0.27]{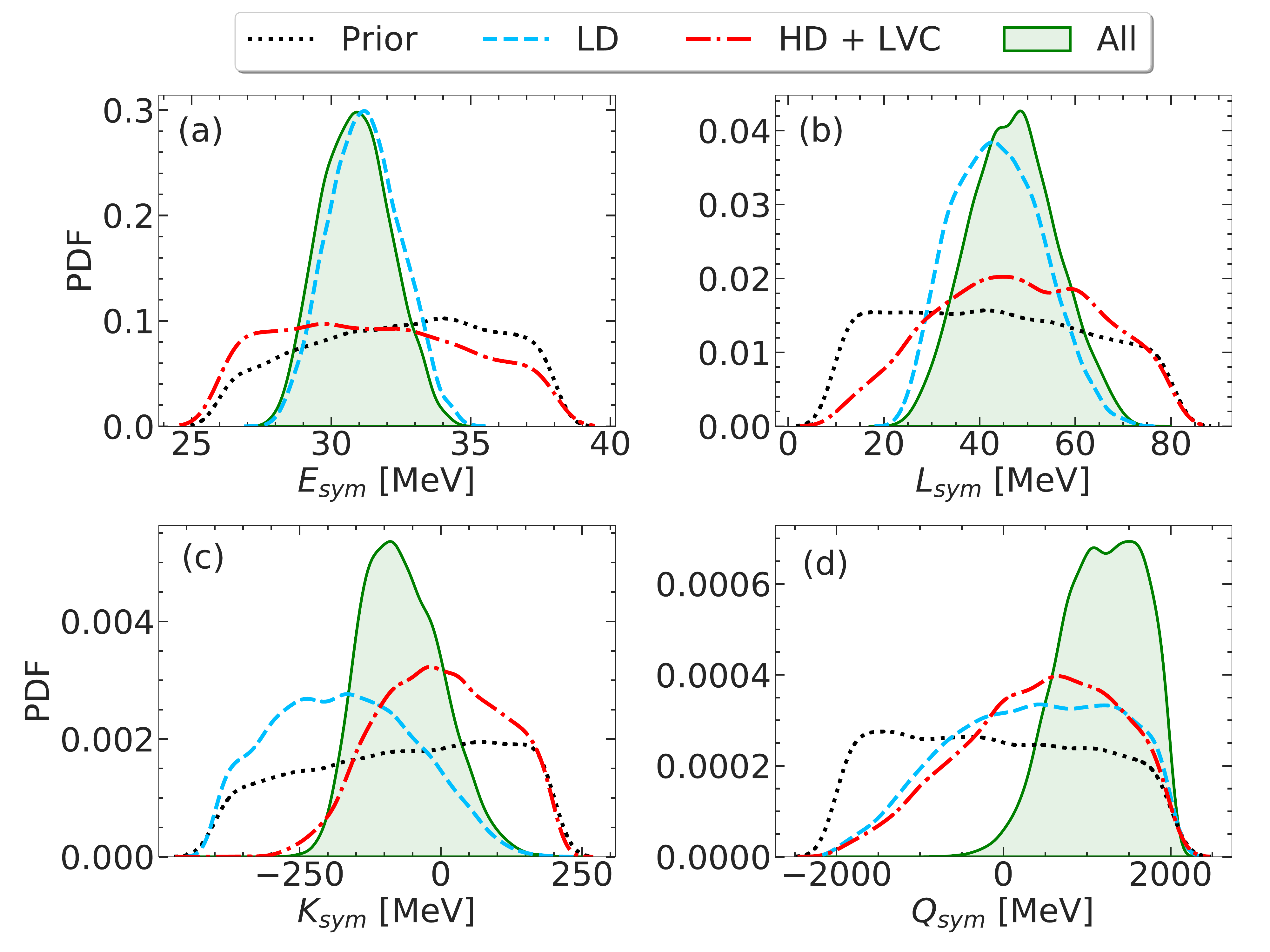}
	\caption{Same as Figure~\ref{isoscalar} but for isovector empirical parameters.  	\label{isovector}}
\end{figure}

In Figures~\ref{isoscalar} and \ref{isovector} we show the probability density distributions (PDFs) of isoscalar and isovector empirical parameters of order $N<4$, respectively. As we have described previously, the distributions labeled as ``Prior" in these figures are not flat, but they carry the information from the experimental nuclear mass measurement. For example, $E_{sat}$, the energy per particle in SNM at saturation, already has a  peaked shape (see  Figure~\ref{isoscalar} (a))  because of this reason. From the HD+LVC distribution in these figures, we can  see that  the astrophysical constraints on NS mass and tidal deformability have almost no effect on low-order parameters. The impact of the chiral EFT filter on the isoscalar parameters of order $N<3$, \textit{i.e.} $E_{sat}$,  $K_{sat}$,  along with $n_{sat}$ is not prominent.  This knowledge can be further reinforced by looking at Table \ref{tab:param-mean-bulk}, where the LD filter hardly improves the constraints on the aforementioned isoscalar parameters.
It can be explained by the fact that the prior intervals of the empirical parameters are chosen based on the current knowledge provided by nuclear physics, in which the deviations of $E_{sat}$, $n_{sat}$, and $K_{sat}$ are already relatively small. 

\begin{table}[H] 
	\centering
	\small
	\setlength{\tabcolsep}{3pt}
	\caption{Medians and $68\%$ confidence limits of EoS empirical parameters of order $N<4$ in the four distributions.  \label{tab:param-mean-bulk}}
	%\tablesize{} % You can specify the fontsize here, e.g., \tablesize{\footnotesize}. If commented out \small will be used.
	\begin{tabular}{lcccccccc}
		\toprule
		&$E_{\rm sat}$  & $n_{\rm sat}$ & $K_{\rm sat}$ & $Q_{\rm sat}$  & $E_{\rm sym}$  &	$L_{\rm sym}$  &$K_{\rm sym}$& $Q_{\rm sym}$  \\
		&[MeV]          & [fm$^{-3}$]   & [MeV]         & [MeV]         & [MeV]         & [MeV]          &   [MeV]          &[MeV]        \\
		\midrule	
		Prior& $-16.25^{+0.61}_{-0.46}$  &$0.159^{+0.008}_{-0.006}$ & $231^{+27}_{-28}$  & $-44^{+693}_{-650}$   &$32.6^{+3.5}_{-3.9}$ & $42^{+24}_{-22}$ & $-62^{+181}_{-210}$ & $-132^{+1394}_{-1290}$\\
		\midrule
		LD& $-15.90^{+0.51}_{-0.50}$  &$0.163^{+0.005}_{-0.008}$ & $239^{+22}_{-30}$   & $-264^{+383}_{-356}$  & $31.2^{+1.3}_{-1.3}$ &$43^{+11}_{-9}$ & $-175^{+136}_{-131}$&$406^{+1026}_{-1116}$  \\
		\midrule
		HD+LVC& $-16.20^{+0.60}_{-0.47}$  &$0.161^{+0.006}_{-0.008}$ &$231^{+27}_{-27}$  & $321^{+467}_{-596}$  & $31.4^{+4.0}_{-3.6}$ &$48^{+18}_{-19}$ & $-2^{+121}_{-113}$&$502^{+891}_{-1054}$  \\
		\midrule
		All& $-15.86^{+0.49}_{-0.50}$  &$0.163^{+0.006}_{-0.007}$ & $249^{+15}_{-23}$  & $-41^{+310}_{-267}$  & $30.9^{+1.3}_{-1.3}$ &$47^{+9}_{-9}$ & $-74^{+78}_{-65}$ & $1207^{+491}_{-539}$ \\
		\bottomrule
	\end{tabular}
\end{table}

Unlike the lower order parameters in the isoscalar sector, the isovector counterparts are quite poorly determined by nuclear physics experiments. As a result, once the constraint from the chiral EFT calculation is included, $E_{sym}$, $L_{sym}$ and  $K_{sym}$ are strongly affected (see Figure~\ref{isovector} and Table \ref{tab:param-mean-bulk}).  
Interestingly, the LD filter also has a non-negligible impact on the high order parameters $Q_{sat}$ and $Q_{sym}$. 
 This is because the chiral EFT calculation gives very precise predictions at very low densities, far from 
	nuclear saturation. In this region,  the high-order parameters  have a non-negligible contribution to the nuclear 
	matter energy. It was shown by Refs. \cite{Hoa2021, Hoaepja2021} that constraining the EoS at very low densities 
$n\sim 0.02-0.1$ fm$^{-3}$ is crucial in studying the crust-core transition.

As one may expect, the constraints from NS observables  (HD+LVC) play an important role on high-order parameters, such as $Q_{sat}$ and $Q_{sym}$, as well as on the poorly constrained isovector compressibility $K_{sym}$. 
One can observe that for these parameters, higher values of the chosen intervals are preferred in the nucleonic hypothesis, having low preference on the softer EoSs.   However, note that this is the net effect of both the radio mass and GW180817  measurements. We have checked that without the constraint on the tidal deformability,  the resulted nuclear matter energies are even higher, which means that the constraint from GW170817 softens the EoS.  

{As  discussed in details in Ref.~\cite{metamodel_1}, the density behavior of realistic functionals can be accurately reproduced up to the central density of massive neutron stars by a Taylor expansion truncated at fourth order, but because of the truncation the parameters of order $N\ge 3$ have to be considered as effective parameters  that govern the high density behavior of the EOS, and do not need to be equal to the corresponding density derivatives at saturation. On the other hand, in the sub-saturation regime, the deviations from the Taylor expansion are accounted for by the low density corrective term that imposes the correct zero density limit~\cite{metamodel_1}. These two effects being completely independent, the meaning of the third and fourth order parameters as  explored by the EFT calculation and the astrophysical observations is not the same, and we can expect that low and high density constraints might point to different values for those parameters. 
	Comparing the dashed and dashed-dotted lines in Figure~\ref{isoscalar} we can   see   that indeed low-density constraints impose lower values of $Q_{sat}$
	with respect to high-density ones.    This means that low energy experiments aimed at a better measurement of $Q_{sat}$ will not improve our empirical knowledge of the high density EOS. Interestingly, the same is not true for $Q_{sym}$, for which the dotted and dash-dotted distributions closely overlap. Even if the present constraints are quite loose, it appears that the skewness of the symmetry energy at saturation $Q_{sym}$ gives a fair description of the behavior of the EOS at high density, while a deviation is observed at the level of the compressibility $K_{sym}$. 
}
We do not   include the results for the fourth-order parameters $Z_{sat}$, $Z_{sym}$, because they have very large uncertainties,  and very little impact from the different constraints. Furthermore, we will see later on that they have almost no correlations to other parameters as well as observables.

\begin{figure}[!htpb]
	\centering
	\includegraphics[scale=0.27]{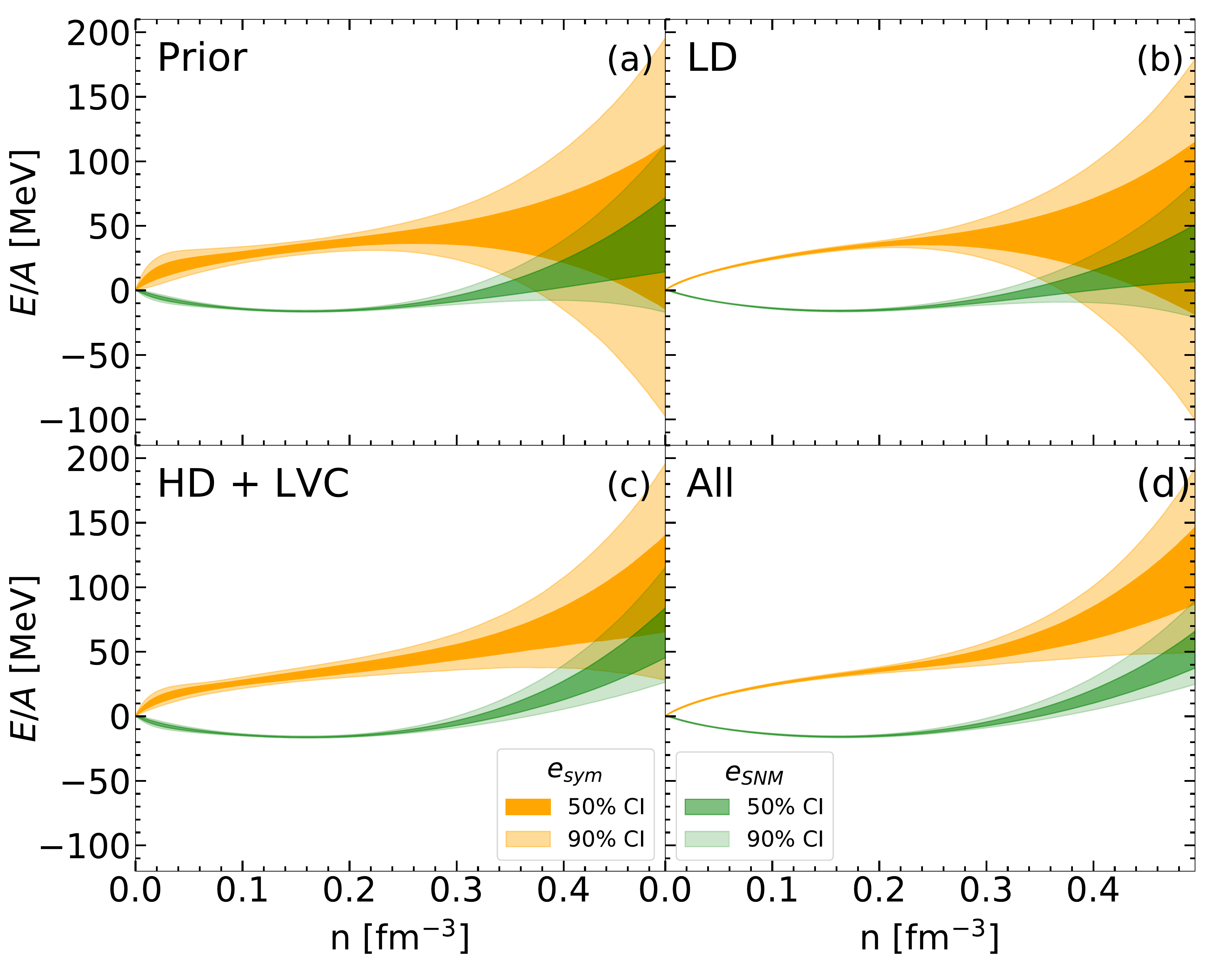}
	\caption{50$\%$ (darker color) and $90\%$ (lighter color) confidence intervals of energy per nucleon of symmetric nuclear matter ($e_{SNM}$, color green) and symmetry energy ($e_{sym}$, color orange) as a function of density $n$.}
	\label{NM_energy}
\end{figure}

In Figure \ref{NM_energy}, we plot the bands for SNM energy per nucleon and symmetry energy at 50\% and 90\% 
confidence intervals for the four posterior distributions as explained in the previous section. The impact of LD and 
HD+LVC filters can be observed by looking at panels (b) and (c) of Figure \ref{NM_energy}, respectively. Their effects 
are appreciated at different density regime, as is also evident from the analysis done in Figures \ref{isoscalar} and 
\ref{isovector} and Table \ref{tab:param-mean-bulk}.

\subsection{ Properties of NS crust}

\begin{table}[H]
	\centering
	\setlength{\tabcolsep}{5pt}
	\caption{Estimations of NS crustal properties in four distributions. The results are presented with medians and $68\%$ confidence limits. \label{tab:crust}  }    
	\begin{tabular}{lcccc}
		\toprule
		&$n_{CC}$ & $P_{CC}$ & $R_{crust}^{1.4}$&$R_{crust}^{2.0}$ \\
		&[fm $^{-3}$]          & [MeV fm$^{-3}$]   & [km]         & [km]     \\
		\midrule
		Prior& $0.087^{+0.033}_{-0.037}$  &$0.163^{+0.281}_{-0.095}$ & $1.13^{+0.25}_{-0.29}$  & $0.706^{+0.165}_{-0.191}$  
		\\
		\midrule
		LD& $0.078^{+0.011}_{-0.011}$  &$0.385^{+0.104}_{-0.097}$ & $1.11^{+0.10}_{-0.14}$   & $0.693^{+0.070}_{-0.079}$  \\
		\midrule
		HD+LVC& $0.079^{+0.023}_{-0.033}$  &$0.141^{+0.202}_{-0.076}$ &$1.05^{+0.20}_{-0.20}$  & $0.627^{+0.126}_{-0.128}$   \\
		\midrule
		All& $0.084^{+0.009}_{-0.010}$  &$0.423^{+0.093}_{-0.090}$ & $1.15^{+0.10}_{-0.08}$  & $0.687^{+0.067}_{-0.067}$   \\
		\bottomrule
	\end{tabular}
\end{table}

In our calculation,  the transition from the solid  crust to the liquid outer core is determined 
by comparing the corresponding energy density of clusterized matter to that of homogeneous matter at $\beta$-equilibrium 
with  the metamodel.  For the crust part, metamodel is extended introducing  surface 
parameters within the compressible liquid drop model (CLDM) approach \cite{CarreauEPJA}. The precision in the prediction 
of the crust-core transition point is crucial in estimating crustal observables, such as crustal mass, thickness, and 
moment of inertia. These quantities are in particular thought to have influence on the origin of the pulsar glitches \cite{glitch}.   In the literature, there are various works devoting to determining the crust-core transition density $n_{CC}$ with different many-body methods and nuclear functionals, spanning a large range of values, such as $n_{CC} = 0.0548$ fm$^{-3}$ in \cite{Grill2012} obtained using Thomas-Fermi calculations for the NL3 functional, or $n_{CC} = 0.081$ fm$^{-3}$ in \cite{Pearson2020} within the full fourth-order extended Thomas-Fermi approach for the BSk24 functional. For this reason, an estimation for the uncertainties of the crustal properties with  Bayesian tools using both the current nuclear physics and astrophysical data, provided by LVC and NICER, are of great importance.

\begin{figure}[!htpb]
	\centering
	\includegraphics[scale=0.3]{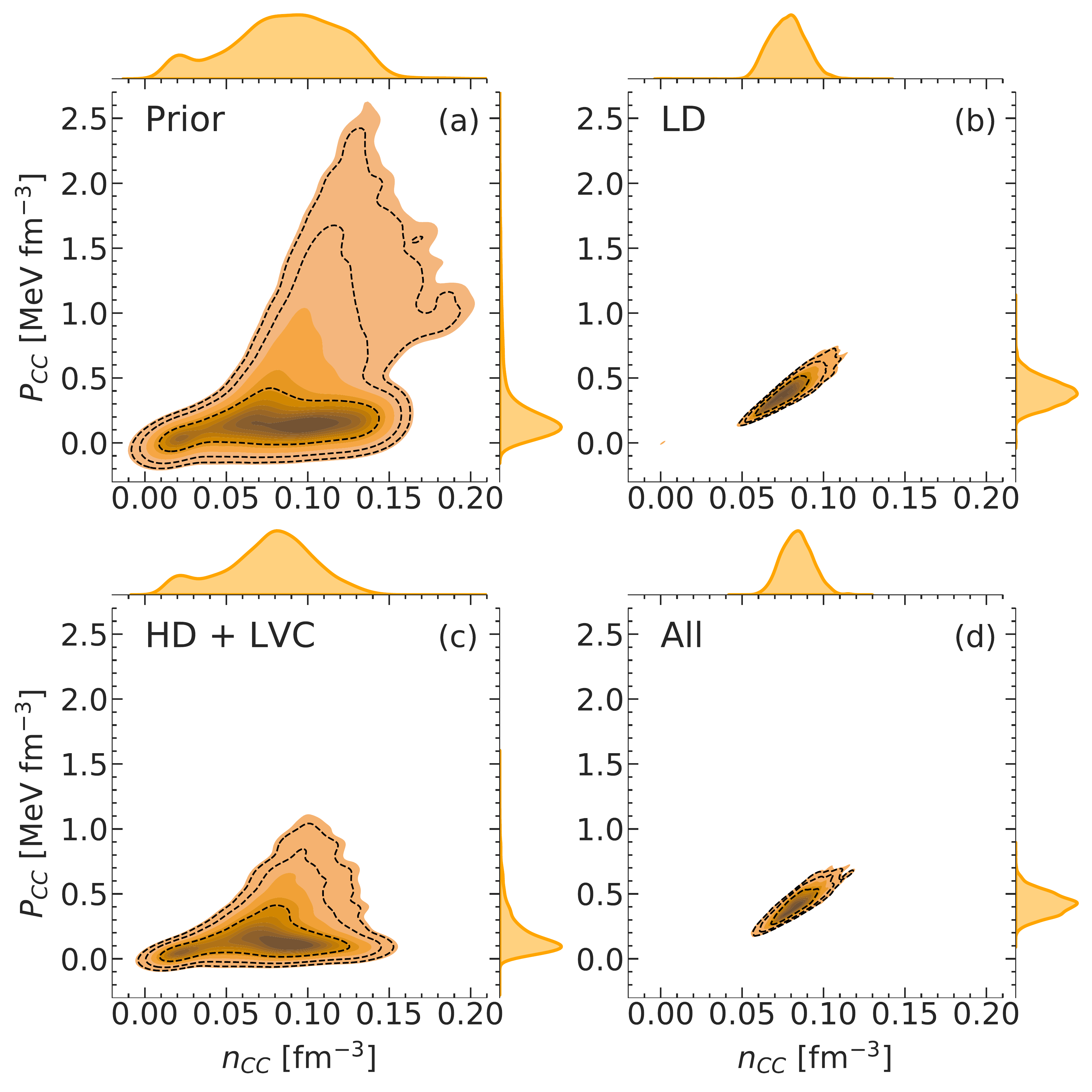}
	\caption{Joint probability density plots of crust-core transition density $n_{CC}$ and pressure $P_{CC}$. The dashed black contours in each panel indicate the 68$\%$, 95$\%$, and $99\%$ confidence regions.}
	\label{CC_transition}
\end{figure}

In Figure~\ref{CC_transition}, we display the joint distributions of the crust-core transition 
density $n_{CC}$ and pressure $P_{CC}$.  The chiral EFT 
calculation plays  an important role in the determination of the crust-core transition point, 
which  is  evident  from  the LD distribution in 
Figure~\ref{CC_transition}  (b).   One can observe that, the 
chiral EFT filter puts stringent limits on both the crust core transition density $n_{CC}$ and pressure 
$P_{CC}$;  very high and very low values of $n_{CC}$ and $P_{CC}$  get 
eliminated.    In  Figure 
\ref{CC_transition} (c) for  the HD+LVC distribution, the most 
noticeable  fact is the  suppression of models  
with  high transition pressures. However, the probability densities of these models are tiny, and they are 
outside of the $95\%$ contour in the prior distribution  (see Figure \ref{CC_transition} (a)). Moreover, they 
are associated to models violating at least one of the following conditions required in the HD+LVC posterior: causality, 
thermodynamics stability,   or  non-negative symmetry energy. In other words, 
the astrophysical constraints on NS maximum mass and tidal deformability have   very 
little  effects on the crust-core transition. This is consistent with our observations for the nuclear matter energy in 
Figure~\ref{NM_energy}  (c) :   the nuclear matter energy in the HD+LVC distribution is not notably constrained at densities
around $n\sim n_{sat}/2$.

\begin{figure}[!htpb]
	\centering
	\includegraphics[scale=0.32]{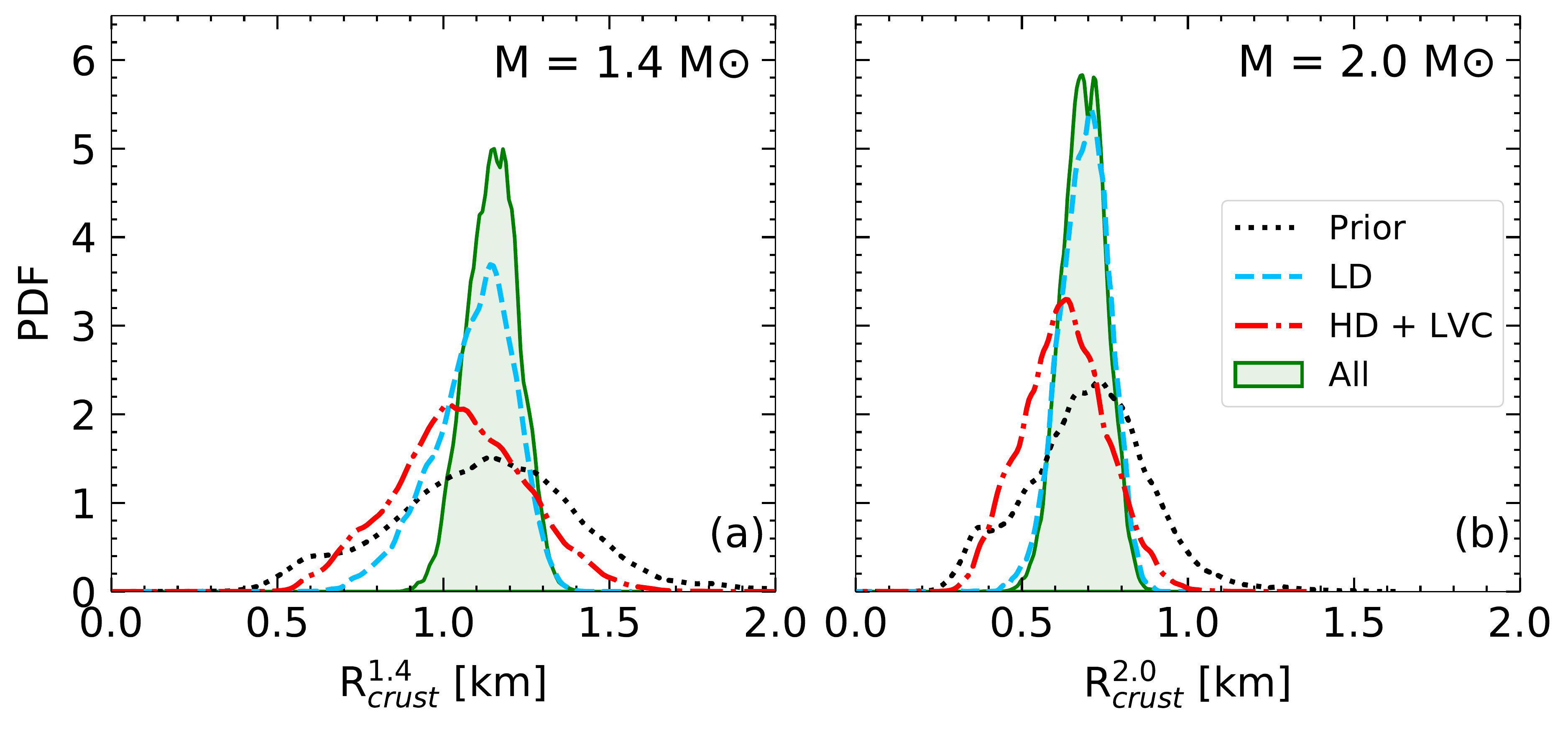}
	\caption{Probability density distributions of crust thickness at $M=1.4 M_{\odot}$ (panel (a)) and $M=2.0 M_{\odot}$ (panel (b)).}
	\label{crust_thickness}
\end{figure}

The crust-core transition point determines astrophysical observables, such as crust thickness, or moment of inertia  \cite{Carreau2019}. In this study, 
we   have chosen  the crust thickness to be the demonstrative quantity. Figure~\ref{crust_thickness} 
presents the  PDFs of  NS crust thicknesses   for  $1.4 M_{\odot}$ and $2.0 M_{\odot}$   NSs. 
In both cases, the uncertainties in the LD distributions are narrowed down compared to the prior, while the effect 
in the HD+LVC distribution is only marginal. This agrees with our conclusions for the crust-core transition
point, that is, the role of the chiral EFT filter is more dominant in the determination of crustal properties. When 
all constraints are taken into account, crust thicknesses  of both $1.4M_{\odot}$ and $2.0M_{\odot}$  NSs 
are known with the relative uncertainties up to $10\%$.  
For a quantitative estimation of the effects of different filters, in Table \ref{tab:crust} we present 
crust-core transition density $n_{CC}$ and pressure $P_{CC}$, along with crustal thickness of $1.4M_{\odot}$ and 
$2.0M_{\odot}$  NSs accompanied by errors on them at 68\% confidence interval. Quite conclusively one can comment 
that the primary effect comes from the LD chiral EFT filter, which also puts stringent constraints when all the 
filters are combined together, denoted as ``All".

\subsection{ NS equation of state}
\begin{figure}[H]
	\centering
	\includegraphics[scale=0.3]{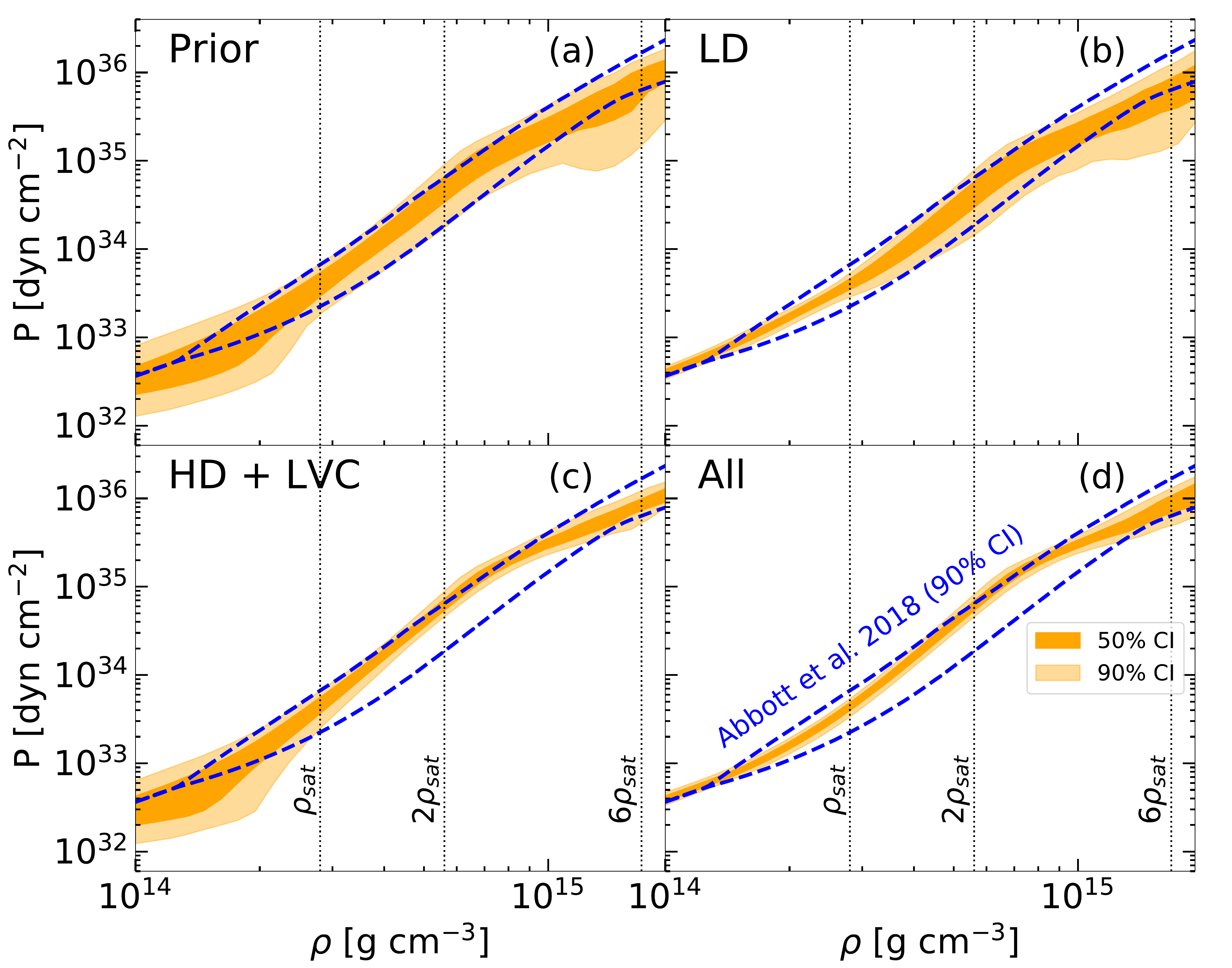}
	\caption{50$\%$ (dark orange) and $90\%$ (light orange) confidence intervals of pressure $P$ as a function of mass density $\rho$ in comparison with the $90\%$ confidence interval of the posterior obtained in Abbott \textit{et al.} 2018 \cite{ligo2}  (blue dashed lines). See text for details.}
	\label{EoS}
\end{figure}

Unlike the crustal properties, HD+LVC filter is expected to   put tighter bounds  on global NS properties, 
which is governed chiefly by the high-density part of 
the EoS. The effects of different filters on the EoS are shown in Figure~\ref{EoS}. The light (dark) orange band 
indicates 90$\%$ (50$\%$) confidence interval. For comparison, we also display the result inferred from the 
gravitational wave data GW170817 by LVC   at $90\%$ level in dashed blue lines   \cite{ligo2}. 
We have also used the same units for mass-density g cm$^{-3}$ as in Ref. \cite{ligo2} for the same reason. In 
this unit the saturation density $n_{sat}$ is denoted as $\rho_{sat}$ ($\simeq 2.8\times10^{14}$ g cm$^{-3}$). 
In  Ref. \cite{ligo2} , Abbott \textit{et al}. have sampled their EoSs at high density using 
the spectral parametrization  \cite{Lindblom2010}. These EoSs are then matched with SLy EoS  \cite{SLy2001} at around   $  \sim\rho_{sat}  /2$.   Incidentally , the authors 
also put some prior criteria similar to our analysis, which are causality, thermodynamic stability, and consistency 
of NS maximum mass with the observation. For the last condition, they put a sharp limit ($M_{max}\geq 1.97M_{\odot}$) 
instead of using a likelihood probability like the one  used in  our 
analysis   (see Eq. \ref{Eq:mmax}). However, we have verified that the difference in the maximum NS treatment 
does not lead to sizable deviation in the final results. 
In Figure \ref{EoS} (a), we can see that our prior distribution perfectly covers the 
whole posterior band   given by GW170817  event \cite{ligo2} with good agreement. 
In our case, the prior distribution carries information from nuclear 
physics experiments and theoretical calculations via the chosen prior intervals of empirical parameters as well as 
the mass fit.  This is why the EoS  in our prior distribution  at 
low densities  is relatively narrow   compared to other analyses. Note that the 
uncertainty   below $\rho_{sat}$,  appears to be  large due to 
the visual effect of the logarithmic scale in the pressure. Once the chiral EFT filter is applied, this uncertainty 
is vastly reduced   (see Figure \ref{EoS} (b)), resulting in a very well-constrained  band and excellently 
compatible with   the posterior constrained by GW170817 data \cite{ligo2} . 
Contrarily, the behavior of the EoS at supra-saturation densities is not constrained by the chiral EFT filter. As a result, a larger dispersion is observed at high densities. 
This dispersion is not as important as in fully agnostic studies \cite{Landry2020}
because of the nucleonic hypothesis that imposes an analytic behavior of the EoS at all densities. This strong hypothesis can be challenged by the astrophysical measurements, and any inconsistency with the observations will reveal the presence of exotic degrees of freedom.

By incorporating the pass-band  
filter $\omega_{HD}$ as well as the condition on the NS maximum mass in Figure \ref{EoS} (c),  the deviation in the lower limit of the pressure at density $\rho \gtrsim 10^{15}$ g cm$^{-3}$ observed in the prior, 
is eliminated.  In particular,  the constraint on the NS maximum mass  sets  a stringent 
limit on the lower bound of the pressure, and posterior EoS is   shifted  significantly 
towards higher values of pressure.    Conversely, the constraint from LVC 
prefers softer EoS, hence setting the limit on the upper bound of the pressure band.   In Figure \ref{EoS} (d), 
when all constraints are combined together, we obtain as expected a narrower band for the EoS than  the one obtained exclusively from GW170817 data \cite{ligo2}. In addition, 
we observe that our EoS is lightly stiffer than the one of   Ref. \cite{ligo2} at around $2-3 \rho_{sat}$. The small width of the EoS and its stiffness are assigned to the semi-agnostic hadronic prior, which represents current nuclear physics knowledge. Nevertheless, the overall agreement is excellent. Thus, it indicates the compatibility of the nucleonic EoS with the gravitational wave GW170817 data.

Comparing   the ``HD+LVC" and ``All" distributions in   Figure~\ref{EoS}  
(c) and (d) , it can be observed that the inclusion of the new NICER measurement does not show any significant 
impact on the EoS.  Similar conclusions have been 
drawn in other studies in the literature. In particular, Pang \textit{et al.} \cite{Pang2021}  has carried out a 
Bayesian analysis using the data from  Riley \textit{et al.} \cite{nicer3} and  Miller \textit{et al.} 
\cite{nicer4}. In both cases ,   they found that the effect 
of the constraint from the radius measurement of PSR J0740+6620 only marginally impact the EoS.    In  
Ref. \cite{Raaijmakers2021} , Raaijmakers \textit{et al.} performed the Bayesian inference with two 
EoS parametrizations: a piece-wise polytropic (PP) model and a  speed-of-sound (CS) model  drawing similar conclusions.   For the 
constraint on  PSR J0740+6620, they employed the data from Riley \textit{et al.} \cite{nicer3}, 
in which the error bar of the radius is  smaller than that obtained in Miller \textit{et al.} \cite{nicer4}. They 
concluded that for the PP models, the impact on the EoS mainly comes from the high mass value of PSR J0740+6620 
because their prior distribution in that mass range is within the 68$\%$ level of the radius measurement (see 
Figure 4 in \cite{Raaijmakers2021}).

\subsection{ Speed of sound in medium}
\begin{figure}[H]
	\centering
	\includegraphics[scale=0.3]{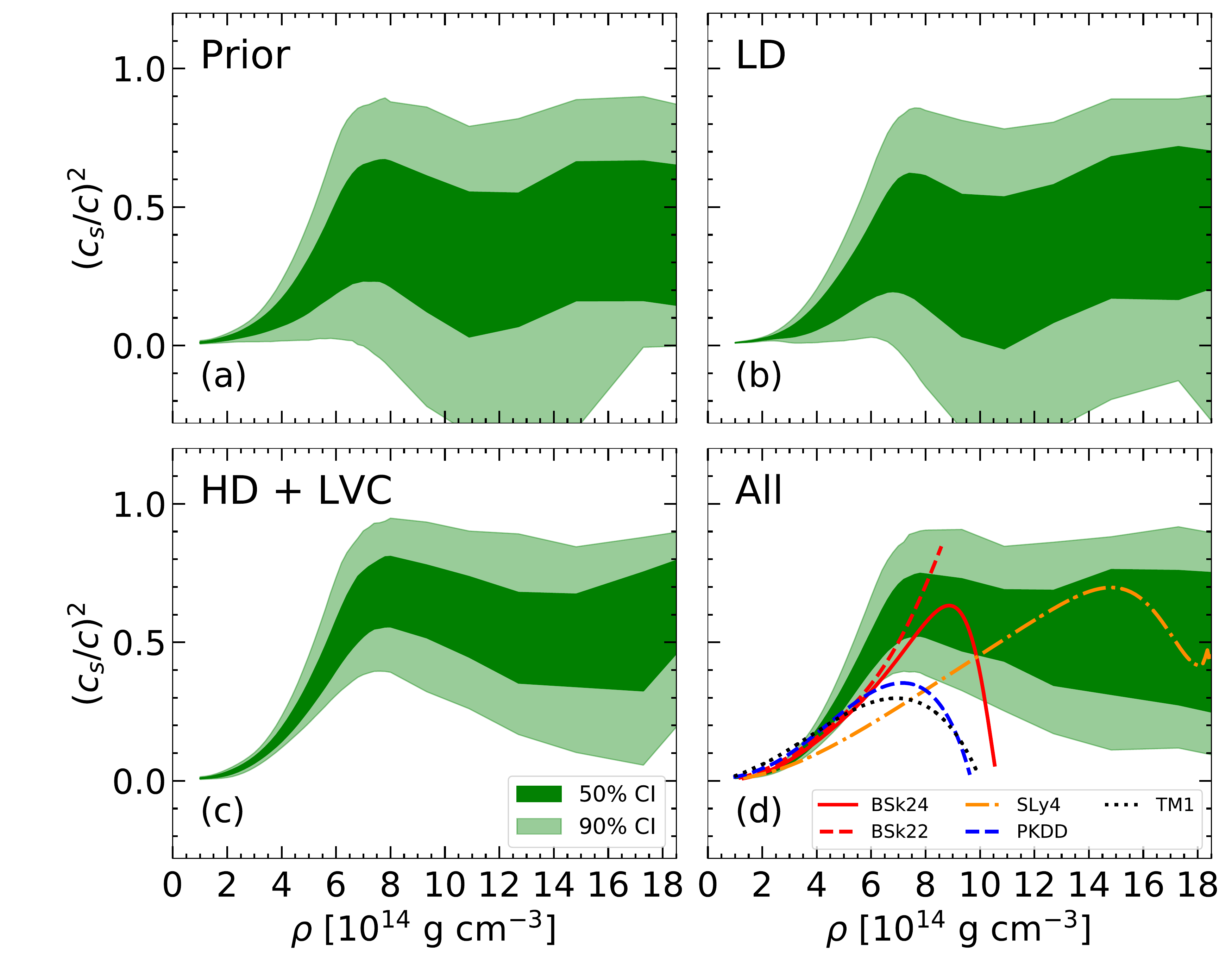}
	\caption{50$\%$ (dark green) and $90\%$ (light green) confidence intervals of sound speed $\left(\frac{c_s}{c}\right)^2$ as a function of mass 
		density $\rho$. Curves in panel (d) show the sound speed of some selected models \cite{Goriely2013,Chabanat1998,Kuwabara1995,Long2004} up to the central density corresponding to the maximum mass. See text for details.}
	\label{sound}
\end{figure}

In Figure \ref{sound}, we plot the velocity of sound in medium as a function of mass density $\rho$ obtained
with four different filters at the 50\% and 90\% confidence intervals, together with the behavior of some selected models \cite{Goriely2013,Chabanat1998,Kuwabara1995,Long2004}. One can observe that for all the filters the
most probable equations of state remain causal up to very high densities ($\sim 6\rho_{sat}$), even though we do not put this
requirement explicitly in our ``Prior" and the ``LD" filters in Figure \ref{sound} (a) and (b), respectively. As expected, 
the behavior of the sound speed is globally structureless. However, we can surprisingly see a trend for a peaked structure, 
which is typically presented in the literature as a signature of a transition to exotic matter. This peak may arise from the 
shoulder observed in Figure~\ref{EoS} above, which is due to the combined constraints of a relatively soft EoS at low density, 
and the requirement of reaching the maximal mass. These conditions lead to a peak in the global distribution, note 
however, not for all models individually (see lines in Figure \ref{sound} (d)). A very small fraction 
	of non-causal models is present due to the fact that we plot the EoS only up to densities where    the nucleon 
	sound velocity is in the interval between 0 and 1. Residual non-causalities (not visible within 90\% confidence 
	interval of Fig. \ref{sound}) originate from the additional lepton contribution in beta-equilibrated matter.
\subsection{ NS observables}
\subsubsection{  Masses and radii }
\begin{figure}[H]
	\centering
	\includegraphics[scale=0.3]{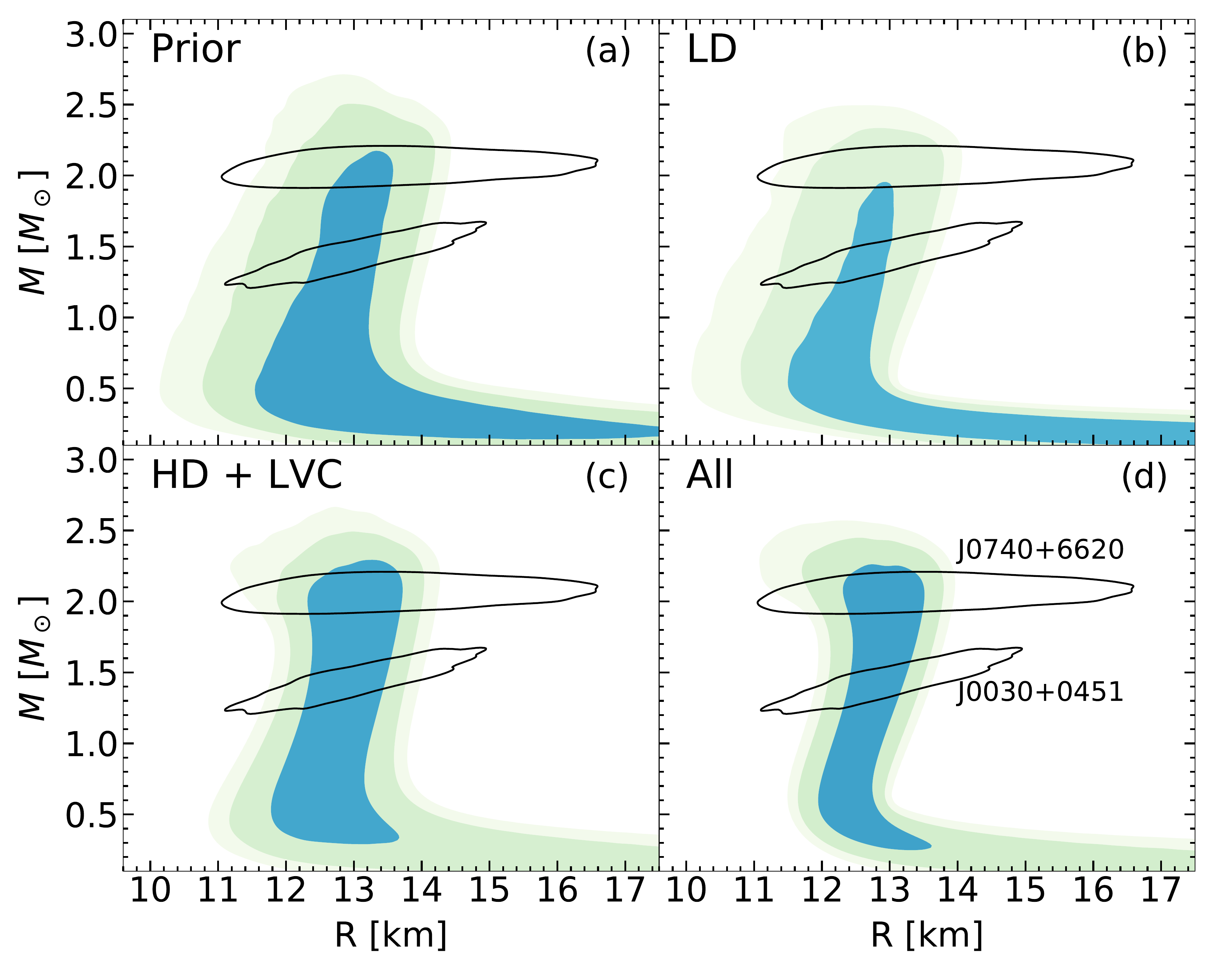}
	\caption{2D density plots of NS mass M as a function of radius R in comparison with two NICER measurements at 68$\%$ (black contours). The three shaded regions in each panel contain 68$\%$, 95$\%$, and 99$\%$  of the distribution. See text for details.}
	\label{MRrelation}
\end{figure}
In Figure 
\ref{MRrelation} we plot for different filters   three shaded regions (from light to dark) sequentially 
containing $99\%$, $95\%$, and $68\%$   confidence intervals for two dimensional 
distribution for mass and radius of NSs. The two black contour lines at low mass and high mass respectively 
indicate $68\%$ of the mass-radius distributions  for PSR J0030+0451   \cite{nicer2} 
and  PSR J0740+6620  \cite{nicer4}. 
One can observe in Figure \ref{MRrelation} (a) that our prior is already quite compatible with both the recent 
NICER observations \cite{nicer2, nicer4}. 
This explains why the effect of the constraints from NICER is globally small in all our distributions. Moreover,
%The   ``LD" filter in Figure \ref{MRrelation}(b) only slightly lowers the compatibility with PSR J0740+6620 data. 
since  in Figure \ref{MRrelation} (c)  we already include the constraint from the radio mass measurement of the 
high-mass pulsar PSR J0348+0432 beforehand, the impact from the mass of PSR J0740+6620  is obscured   in 
Figure \ref{MRrelation} (d). Additionally, the large uncertainty in the new radius measurement does not help constrain further the EoS. 
The compatibility of NICER measurement 
and our distributions implies that a  nucleonic EoS is  flexible  enough 
to reproduce those dense-matter observations. In Ref. \cite{Pang2021}, Pang \textit{et al.} computed the Bayes 
factor to study the possibility of having a strong first-order  phase  transition from 
nuclear to  quark  matter in NS. If the data from Miller \textit{et al.} \cite{nicer4} is used, 
the Bayes factor changes from 0.265 to 0.205. Even though the effect from PSR J0740+6620 is not significant, 
a  decrease in the  Bayes factor   
points to the fact  that a first-order phase transition  to quark matter  is disfavored. 
Similarly, Legred \textit{et al.} \cite{Legred2021} found that the Bayes factor for EoSs having one stable branch 
against those with at least one disconnected hybrid star branch is 0.156 (0.220) with (without) the PSR J0740+6620 
measurement.  Both these studies  censure  the possibility of  a strong phase 
transition  and support  the suitability of the hadronic EoS with respect to NS observables, which is in line with our present analysis.

\begin{figure}[!htpb]
	\centering
	\includegraphics[scale=0.34]{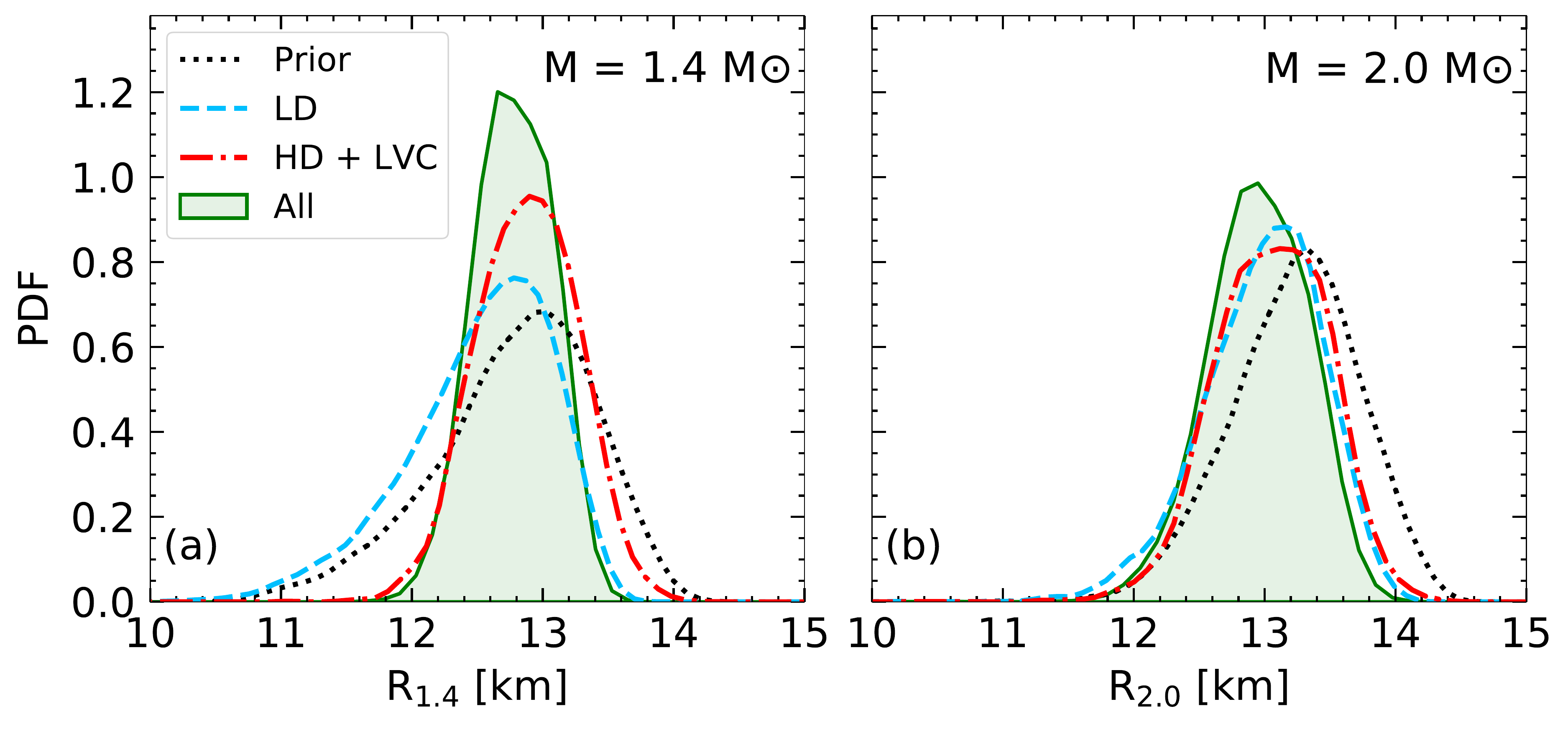}
	\caption{Probability density distributions of NS radii at $M=1.4 M_{\odot}$ (panel (a)) and $M=2.0 M_{\odot}$ (panel (b)).}
	\label{radius}
\end{figure}

Figure~\ref{radius} 
displays the marginalized distributions of NS radii, $R_{1.4}$ and $R_{2.0}$, of the canonical mass 1.4${M_\odot}$ 
(panel (a)) and the typical high mass $2.0M_{\odot}$ (panel (b)) , respectively. The dashed blue lines represent 
the PDFs obtained when chiral EFT   (LD) filter is applied. We can see from the  figure that this filter 
puts a constraint on the  upper  bound of the distributions.   It rejects models with $R_{1.4} \gtrsim 13.6$ km and $R_{2.0} \gtrsim 14.0$ km. In the HD+LVC distribution 
for $1.4M_{\odot}$ NS, the constraint from GW170817 softens the EoS, hence constraining the upper bound of $R_{1.4}$, 
while the requirement on the NS maximum mass filters out very soft EoSs, which put a limit to the lower bound of 
$R_{1.4}$. As a result, these two competing effects provide us with a relatively   narrow  range 
on the radius. In particular, $R_{1.4}\in \left[11.8,14.0\right]$ km (see 
red dashed-dotted line in panel (a)). In the case of $R_{2.0}$, the constraint from radio mass measurement of 
PSR J0348+0432 becomes redundant because all distributions must support $2.0M_{\odot}$ 
NS resulting in  no effect shown in the lower value of $R_{2.0}$. Therefore, in the HD+LVC distribution of 
$R_{2.0}$, the constraint only comes from the LVC measurement. Furthermore, this figure also tells us that the 
impacts on $R_{2.0}$ from the gravitational signal GW170817 and chiral EFT calculation are very similar, even though 
they affect two different regions of the EoS. Specifically, the former controls the EoS in the NS core, hence the 
core radius, while the latter dominates the crust EoS, hence the crust thickness. The prediction in the form of median 
and $68\%$ credible limits for $R_{1.4}$ ($R_{2.0}$) when all constraints are applied together is 12.78$^{+0.30}_{-0.29}$ 
(12.96$^{+0.38}_{-0.37}$) km. In Miller \textit{et al.} \cite{nicer4}, the authors employed three EoS models,  namely Gaussian, spectral, and PP. The values of $R_{1.4}$ for these three models are respectively $12.63^{+0.48}_{-0.46}$ 
km, $12.30^{+0.54}_{-0.51}$ km, and $12.56^{+0.45}_{-0.40}$ km   at 68\% confidence limit. Despite the difference 
in EoS sampling methods, these results are in excellent agreement with   the results obtained in 
the present work. Using also the likelihood from PREX-II measurement of $R_{skin}^{208}$ \cite{Adhikari2021}, 
Ref.  \cite{Biswas2021} obtains $R_{1.4} = 12.61^{+0.36}_{-0.41}$ km, which is also consistent with our prediction. 

\begin{figure}[H]
	\centering
	\includegraphics[scale=0.32]{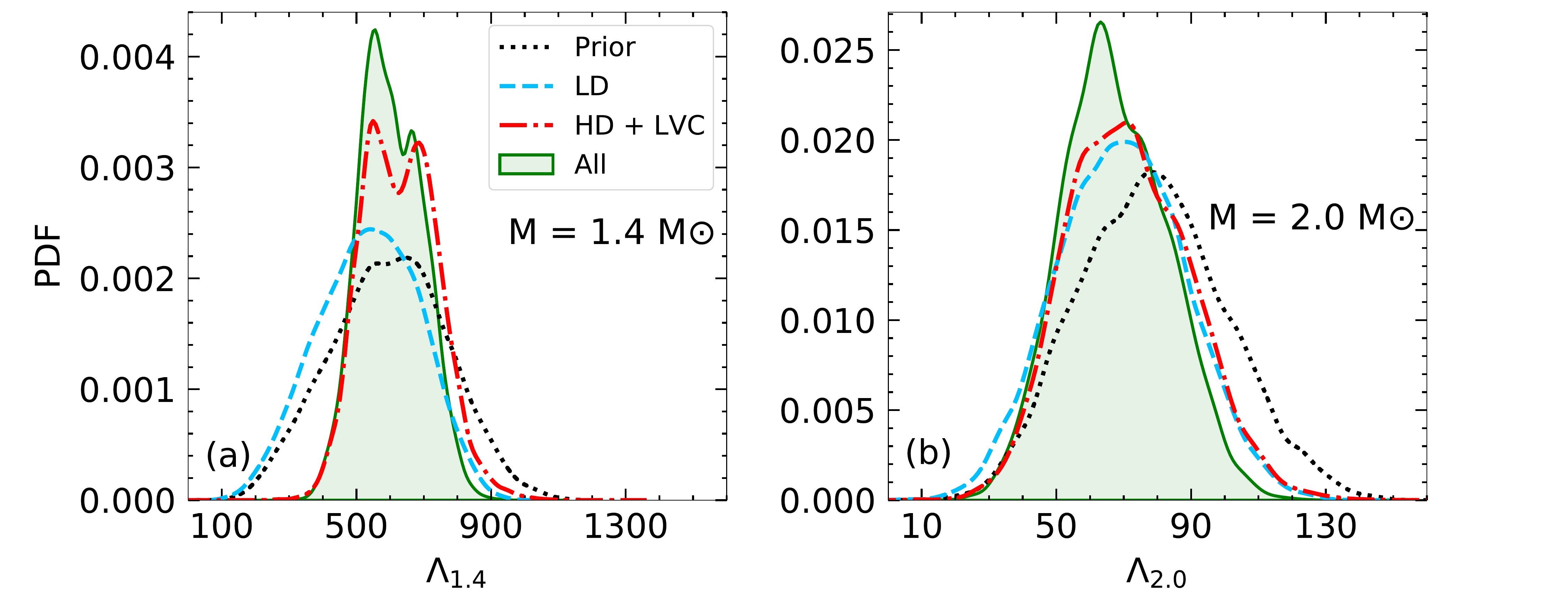}
	\caption{Probability density distributions of NS dimensionless tidal deformabilities at $M=1.4 M_{\odot}$ (panel (a)) and $M=2.0 M_{\odot}$ (panel (b)).}
	\label{lambda}
\end{figure}

The  dimensionless tidal deformability $\Lambda$  in Eq. \ref{eq:lambda} suggests a relation 
between $\Lambda$ and $R$   for a NS of given  mass $M$. However, this relationship 
is not straightforward due to the complex radius dependence of the tidal Love number $k_2$  \cite{Hinderer2008, 
	Binnington2009, Damour2009}. The relation between $R$ and $\Lambda$ , particularly for the mass $M=1.4 M_{\odot}$ 
has been investigated in several works \cite{Malik2018, Fattoyev,Annala2018,Lourenco2019}. Interestingly, 
Figure~\ref{lambda} shows that the distributions of $\Lambda_{1.4}$ and $\Lambda_{2.0}$ behave in accordance with the  
corresponding radius distributions in Figure~\ref{radius}. This may indicate a strong positive correlation between 
these two quantities, which will be discussed later on. In addition, we estimate the $90\%$ confidence boundaries 
of $\Lambda_{1.4}$ ($\Lambda_{2.0}$) to be $\Lambda_{1.4} \in [463,757]$  ($\Lambda_{2.0} \in [43,94]$). This 
prediction of $\Lambda_{1.4}$ agrees excellently with the upper bound extracted from GW170817 signal in Ref. 
\cite{ligo1} using TaylorF2 model, that is, $\Lambda_{1.4} \le 800$. The limit in $\Lambda_{1.4}$ has been improved 
in Ref. \cite{ligo2}, in which the more realistic waveform PhenomPNRT was employed, and they obtained $\Lambda_{1.4} 
\in [70,580] $ at $90 \%$ confidence level for the EoS-insensitive analysis  \cite{ligo1, ligo2}.  Our distribution is still compatible with this result, 
but it suggests a slightly too stiff EoS
in the nucleonic hypothesis.

%\subsection{ \sout{Central proton fraction}}
\subsubsection{Composition}
The determination of the proton fraction is crucial for studying NS cooling. The most efficient cooling mechanism of 
NS is through the  direct Urca (dUrca) neutrino emission process. This process is described by  the successive following reactions:
% If the documentclass option "submit" is chosen, please insert a blank line before and after any math environment (equation and eqnarray environments). This ensures correct linenumbering. The blank line should be removed when the documentclass option is changed to "accept" because the text following an equation should not be a new paragraph.

\begin{align}
	&n  \rightarrow p + l + \bar{\nu}_l\\
	&p + l  \rightarrow n + \nu_l,
\end{align}
where $l=\{e^{-},\mu^{-}\}$. From the momentum and charge conservations, one can derive the expression for the threshold, below which the dUrca process is forbidden:
%If the documentclass option "submit" is chosen, please insert a blank line before and after any math environment (equation and eqnarray environments). This ensures correct linenumbering. The blank line should be removed when the documentclass option is changed to "accept" because the text following an equation should not be a new paragraph.

\begin{equation}
	x_{DU} =\frac{1}{1+(1+x_{ep}^{1/3})^3},
\end{equation}
where $x_{ep} (=x_e/x_p) $ is the ratio   between electron   and proton fraction. Values of $x_{DU}$ can vary in 
the range from $x_{DU} \simeq  1/9 $ in the case of no muons ($x_{ep} =1$) to $x_{DU} 
\simeq   0.148 $ at the limit of massless muons ($x_{ep} =0.5$) \cite{Lattimer1991,Klahn2006}.

\begin{figure}[!htpb]
	\centering
	\includegraphics[scale=0.35]{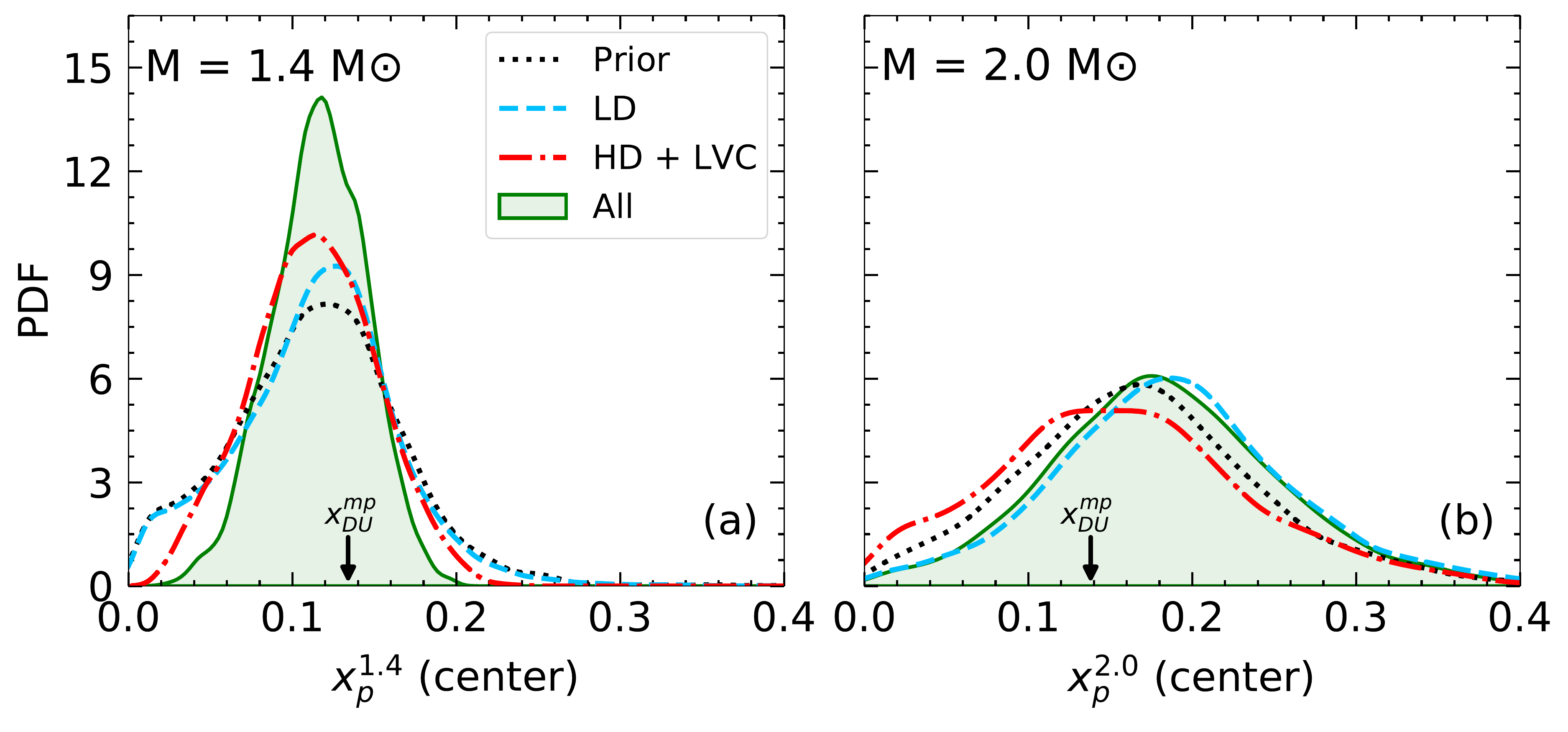}
	\caption{Probability density distributions of central proton fractions of NS at M=1.4 $M_{\odot}$ and M=2.0 $M_{\odot}$. The arrow in each panel indicates the most probable value of $x_{DU}$. Panel (a): $x_{DU}^{mp} \simeq 0.134$. Panel (b): $x_{DU}^{mp} \simeq 0.138$. At each value of mass, value of $x_{DU}^{mp}$ are very similar in four distributions. See text for details.} 
	\label{proton_fraction}
\end{figure}

Figure~\ref{proton_fraction} shows the PDFs of proton fractions calculated at the center of NS  
with  $M=1.4 M_{\odot}$ and $M=2.0 M_{\odot}$. The black arrow in each panel indicates the most probable value of 
$x_{DU}$, calculated for the central density, denoted as $x^{mp}_{DU}$ . We find that this quantity is independent of the constraint used. 
Furthermore, $x^{mp}_{DU}$ only depends weakly on  NS mass,  $x^{mp}_{DU} \simeq 0.134$ (0.138) for $M=1.4$ (2.0) 
$M_{\odot}$. For both masses, the distributions of $x_p$ extend to higher values than the corresponding 
threshold   $x^{mp}_{DU}$ . Therefore, it is possible for the dUrca process to operate even in 
NS of mass $1.4M_{\odot}$. Nevertheless, this fast cooling channel is more likely to happen in  heavier  NSs  due to the higher median and deviation of the $x_p$ distribution. 
By integrating the PDF to find the area under the curve for $x_p \geq x^{mp}_{DU}$, we estimate the possibility for 
the dUrca process in NS of mass $1.4M_{\odot}$ ($2.0M_{\odot}$)  to be approximately $26\%$ (72$\%$).  For 
a more definitive evaluation,  the predictions of NS central proton fractions   along with 
the radius and tidal deformability for NSs of mass $1.4M_{\odot}$ and $2.0M_{\odot}$ at 68\% confidence limit 
are listed in Table \ref{tab:observables} .

\begin{table}[H]
	\centering
	\setlength{\tabcolsep}{5pt}
	\caption{Medians and $68\%$ confidence intervals of NS radii, dimensionless tidal deformabilities, and central proton fractions at $M=1.4 M_{\odot}$ and $M=2.0 M_{\odot}$. \label{tab:observables}  }    
	\begin{tabular}{lcccccc}
		\toprule
		&$R_{1.4}$  &$R_{2.0}$&  $\Lambda_{1.4}$ & $\Lambda_{2.0}$ &	$x_p^{1.4}$  & $x_p^{2.0}$ \\
		&[km]         & [km]          &         &   & &    \\
		\midrule
		Prior &$12.85^{+0.52}_{-0.69}$ & $13.26^{+0.45}_{-0.52}$
		& $601^{+171}_{-182}$ & $78^{+23}_{-22}$
		&$0.115^{+0.047}_{-0.052}$
		&$0.166^{+0.073}_{-0.070}$
		\\
		\midrule
		LD& $12.61^{+0.45}_{-0.64}$
		&$13.03^{+0.39}_{-0.49}$ & $541^{+151}_{-162}$
		&$70^{+19}_{-19}$  &
		$0.117^{+0.041}_{-0.052}$ 
		&$0.187^{+0.072}_{-0.067}$ \\
		\midrule
		HD+LVC& $12.89^{+0.38}_{-0.40}$ &$13.07^{+0.42}_{-0.44}$ & $626^{+114}_{-107}$
		&$71^{+20}_{-17}$ 
		&$0.113^{+0.038}_{-0.039}$
		&$0.154^{+0.079}_{-0.074}$ \\
		\midrule
		All& $12.78^{+0.30}_{-0.29}$ &$12.96^{+0.38}_{-0.37}$ & $598^{+105}_{-85}$
		& $66^{+18}_{-14}$ 
		&$0.117^{+0.027}_{-0.030}$ 
		&$0.181^{+0.070}_{-0.065}$ \\
		\bottomrule
	\end{tabular}
\end{table}

\subsection{ Pearson correlations}

\begin{figure}[!htpb]
	\centering
	\includegraphics[scale=0.4]{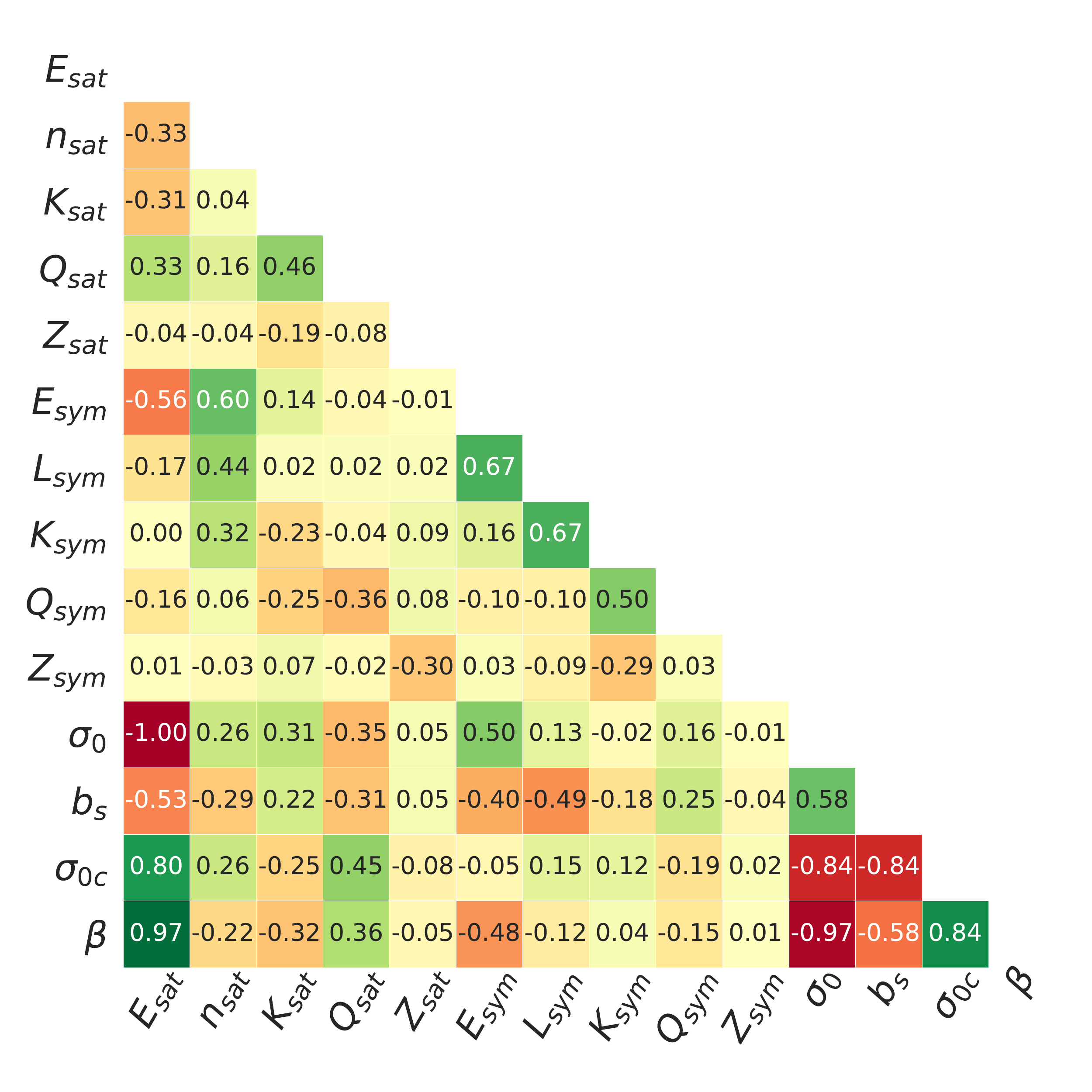}
	\caption{Pearson correlation matrix among  bulk and surface empirical parameters in the case all filters are applied.  } %Pearson correlation matrix among some observables in the case all filters are applied.}
	\label{corr_params}
\end{figure}

\begin{figure}[!htpb]
	\centering
	\includegraphics[scale=0.5]{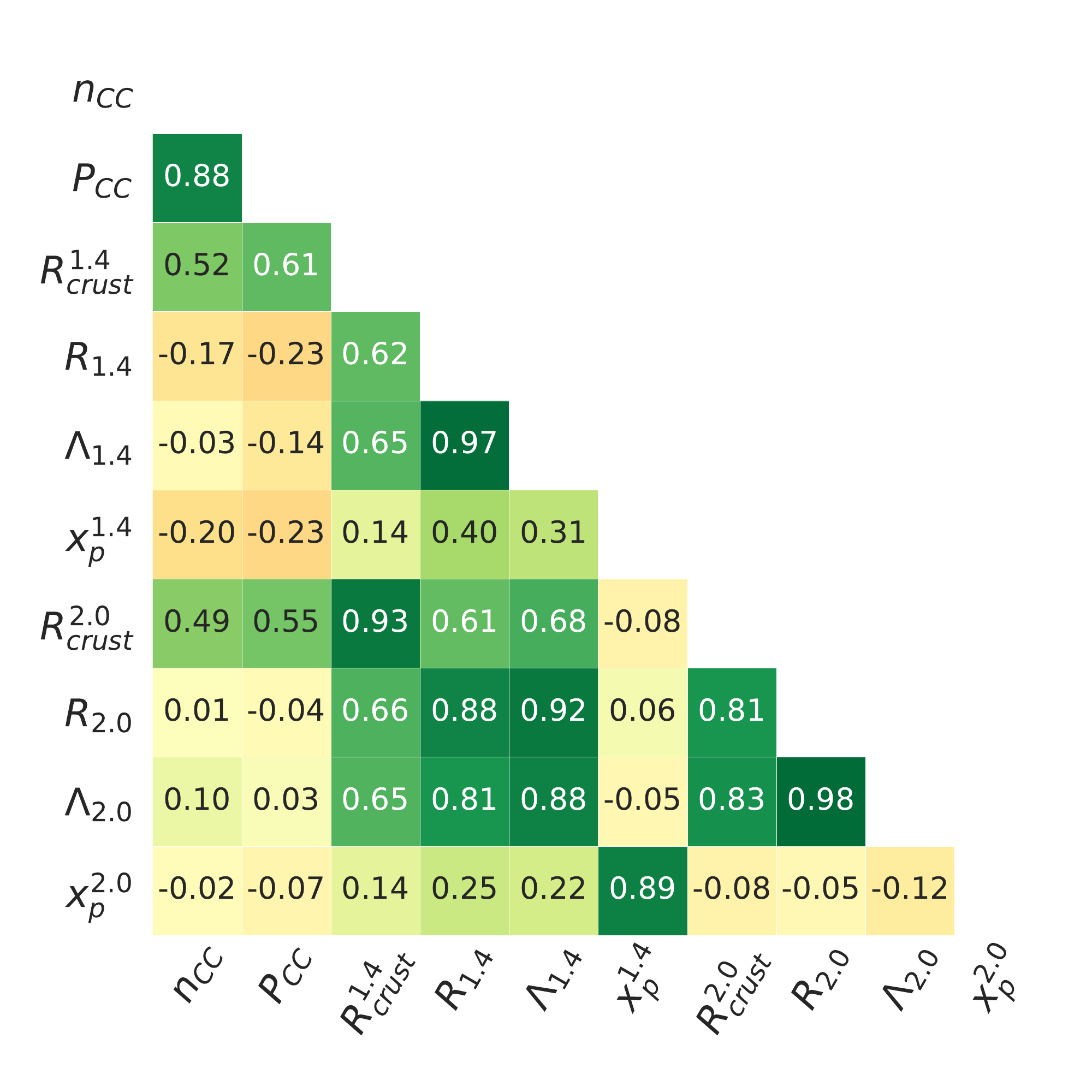}
	\caption{Pearson correlation matrix among some observables in the case all filters are applied.}
	\label{corr_obs}
\end{figure}

%\sout{The linear correlation between two variables ($x$, $y$) can be evaluated using Pearson correlation coefficient as} 
Studying correlations among parameters and observables reveals a great deal about the many facets of 
multi-parametric model calculations \cite{Dobaczewski14}. The most frequently employed tool for this purpose 
is the linear Pearson 
correlation, which is defined for two quantities $x$ and $y$ ($x,y$ can be parameters of the model or any observable 
calculated from it) as, 

\begin{equation}
	corr({x,y})=\frac{cov(x,y)}{\sigma_x\sigma_y},
\end{equation}
where $cov (x,y)$ is the covariance  between $x$ and $y$ , and $\sigma_x$ ($\sigma_y$) is the standard 
deviation on $x$ ($y$). 

Figure~\ref{corr_params} displays the   Pearson  correlation  coefficients  among all bulk, 
surface, and curvature parameters in the case   where all constraints are applied. Since the bulk parameters 
are initially by construction uncorrelated in the flat prior distribution, we can easily assign the induced correlations to the 
different filters   employed. It is shown in the figure that there is a perfect 
negative correlation between the surface tension of symmetric matter $\sigma_0$ and the saturation energy $E_{sat}$, 
with $corr({\sigma_0,E_{sat}}) = -1$. A similar result   was found in   Ref. 
\cite{CarreauEPJA}. The parameters associated to the curvature ($\sigma_{0c}$ and $\beta$), 
on the other hand, exhibit strong positive   correlations  with $E_{sat}$.  These 
correlations  appear due  to 
the fit of the surface and curvature parameters to the experimental 
nuclear mass table. In addition, if the prior is only constrained by   the 
experimental masses of nuclei, we also find a strong correlation between $b_s$ and $E_{sym}$, which are the two 
main parameters governing the energy of asymmetric nuclear matter. However, once the filter from chiral EFT 
calculation is applied, $E_{sym}$ is tightly constrained, and hence   the correlation   get 
blurred. Similar to Refs. \cite{CarreauEPJA, metamodel_2}, no significant correlations are found to be induced 
by the astrophysical constraints. The correlations among the bulk parameters shown in Figure~\ref{corr_params} 
are resulted from the chiral EFT constraint. In particular, the symmetry energy $E_{sym}$ has a moderate 
(anti)correlation with ($E_{sat}$) $n_{sat}$. Stronger correlations are found among the isovector parameters, 
which are $corr(E_{sym}, L_{sym}) = 0.67$ and $corr(L_{sym},K_{sym}) = 0.67$. The former is found in several 
works (see Refs. \cite{Lim2019,CarreauEPJA,Margueon-Gulminelli2019} and references therein for a review), and 
the latter is also studied in Refs. \cite{Danielewicz2009, Chen2009, Vidana2009, Ducoin2010, Santos2014, 
	Mondal17, Mondal18}. 
Slight correlations between high-order parameters, $K_{sat}-Q_{sat}$ and $K_{sym}-Q_{sym}$, are also induced 
due to the narrow EFT energy bands at very low densities.
\begin{figure}[htbp]
	\centering
	\includegraphics[scale=0.38]{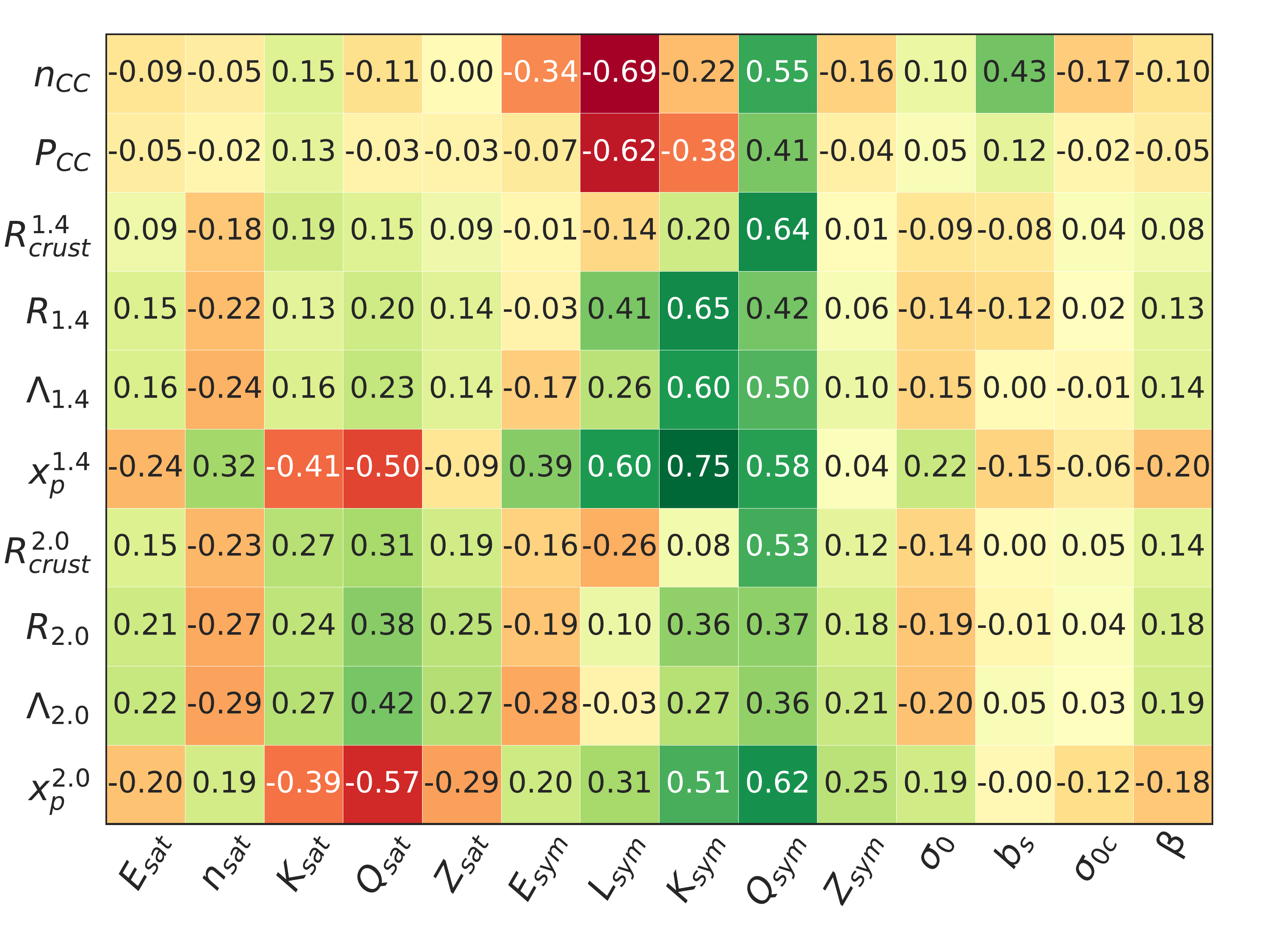}
	\caption{Pearson correlation coefficients between some observables with the empirical and surface parameters in the case all filters are applied.}
	\label{corr_params_obs}
\end{figure}

%\sout{Interesting c}
Correlations  among   different  observables 
found in our study are  plotted in Figure~\ref{corr_obs}. The strongest correlations in this 
matrix are the well known ones between radius and dimensionless tidal deformability, $corr(R_{1.4}, \Lambda_{1.4})=0.97$ 
and $corr(R_{2.0}, \Lambda_{2.0})=0.98$. This explains  the similarity in the distributions of $R$ 
and $\Lambda$ seen in Figures~\ref{radius} and \ref{lambda}. There is also a strong positive correlation between 
$n_{CC}$ and $P_{CC}$. This correlation is visible also in the joint distribution plot in Figure~\ref{CC_transition}. 
We have mentioned before  that  the determination of the transition point from crust 
to core is important in predicting crustal observables; this is again confirmed by the 
correlation coefficients between the crust thickness with the transition density and pressure.   
Finally, the correlations between the observables and parameters are also computed. The correlation matrix is 
shown in Figure~\ref{corr_params_obs}. For most of the cases, the most influential parameters are from the 
isovector channel, which are $L_{sym}$, $K_{sym}$, and $Q_{sym}$. The only exception is for the proton fraction, 
where high-order isoscalar parameters have negative correlations. 
This correlation study clearly demonstrates that astrophysical observables have some marginal influence 
on the higher order nuclear matter properties, which points towards two conclusions: (a) the low density nuclear 
physics data have big influence on constraining the lower order parameters; (b) we need more precise astrophysical 
data to tighten the constraints on higher order parameters.    Conversely, to have 
more accurate prediction  on    astrophysical properties, we need to reduce the 
uncertainties in these   higher-order  parameters   from other sources, \textit{e.g.}, heavy ion collisions \cite{HADES}.  
\section{Conclusions}
%
%This section is not mandatory, but can be added to the manuscript if the discussion is unusually long or complex.
To conclude, we have jointly analyzed different constraints on the nuclear matter EoS coming from nuclear experiments, ab-initio nuclear theory, and several new astrophysical observational data, including the very recent simultaneous observation of mass-radius of PSR J0348+0432 and PSR 
J0740+6620 from NICER collaboration as well as LIGO-Virgo observation of tidal deformability in GW170817 event. Imposing all these different constraints in a Bayesian framework, we have challenged the hypothesis of a fully analytical (continuous and derivable at all orders) EoS, as obtained in the case where dense baryonic matter is purely constituted of neutrons and protons without any phase transition or exotic degrees of freedom. %{\red What about plotting also the speed of sound, to show the specificity of the nucleonic hypothesis?}

Particularly, we have observed that if we have a nuclear physics informed prior 
including the binding energy data of the whole nuclear chart and chiral EFT constraints on low density SNM and PNM, 
the posterior for mass-radius of NSs are already in line with NICER observations. Contrarily, bounds on high density 
matter from radio astronomy observation of NS of 2 solar mass and GW170817 data on tidal deformability is reasonably 
well appreciated. With the present knowledge on astrophysical observations, we predict that the direct Urca cooling 
is possible with  non-negligible probability (27\%) even in a NS with mass as low as 1.4$M_{\odot}$, which increases 
much further $\sim 72\%$ for a NS of 2.0$M_{\odot}$. This might also be very crucial to (in)validate the nucleonic 
hypothesis of high density matter. As all current data on astrophysical observations comply with the nucleonic 
hypothesis within our metamodel approach, we need much more stringent constraints from observations to conclusively 
establish (reject) the presence of exotic degrees of freedom in high density matter.
\section*{Acknowledgment}
 The authors acknowledge partial support from the IN2P3 Master Project “NewMAC".


\begin{thebibliography}{999}
	% Reference 1	
	%\bibitem[Author1(year)]{ref-journal}
	%Author~1, T. The title of the cited article. {\em Journal Abbreviation} {\bf 2008}, {\em 10}, 142--149.
	% Reference 2
	%\bibitem[Author2(year)]{ref-book1}
	%Author~2, L. The title of the cited contribution. In {\em The Book Title}; Editor1, F., Editor2, A., Eds.; Publishing House: City, Country, 2007; pp. 32--58.
	% Reference 3
	%\bibitem[Author3(year)]{ref-book2}
	%Author 1, A.; Author 2, B. \textit{Book Title}, 3rd ed.; Publisher: Publisher Location, Country, 2008; pp. 154--196.
	% Reference 4
	%\bibitem[Author4(year)]{ref-unpublish}
	%Author 1, A.B.; Author 2, C. Title of Unpublished Work. \textit{Abbreviated Journal Name} stage of publication (under review; accepted; in~press).
	% Reference 5
	%\bibitem[Author5(year)]{ref-communication}
	%Author 1, A.B. (University, City, State, Country); Author 2, C. (Institute, City, State, Country). Personal communication, 2012.
	% Reference 6
	%\bibitem[Author6(year)]{ref-proceeding}
	%Author 1, A.B.; Author 2, C.D.; Author 3, E.F. Title of Presentation. In Title of the Collected Work (if available), Proceedings of the Name of the Conference, Location of Conference, Country, Date of Conference; Editor 1, Editor 2, Eds. (if available); Publisher: City, Country, Year (if available); Abstract Number (optional), Pagination (optional).
	% Reference 7
	%\bibitem[Author7(year)]{ref-thesis}
	%Author 1, A.B. Title of Thesis. Level of Thesis, Degree-Granting University, Location of University, Date of Completion.
	% Reference 8
	%\bibitem[Author8(year)]{ref-url}
	%Title of Site. Available online: URL (accessed on Day Month Year).	
	\bibitem{nicer1} Riley, T. E.; Watts, A. L.; Bogdanov, S.; Ray, P. S.; Ludlam, R. M.; Guillot, S.; Arzoumanian, Z.; Baker, C. L.; Bilous, A. V.; Chakrabarty, D.; et al. A NICER View of PSR J0030+0451: Millisecond Pulsar Parameter Estimation.  {\em Astrophys. J. Lett.} {\bf 2019}, {\em 887}, L21, DOI: 10.3847/2041-8213/ab481c.
	
	\bibitem{nicer2} Miller, M. C.; Lamb, F. K.; Dittmann, A. J.; Bogdanov, S.; Arzoumanian, Z.; Gendreau, K. C.; Guillot, S.; Harding, A. K.; Ho, W. C. G.; Lattimer, J. M.; et al. PSR J0030+0451 Mass and Radius from NICER Data and Implications for the Properties of Neutron Star Matter. {\em  Astrophys. J. Lett.} {\bf 2019}, {\em 887}, L24, DOI: 10.3847/2041-8213/ab50c5.
	
	\bibitem{nicer3} Riley, T. E.; Watts, A. L.; Ray, P. S.; Bogdanov, S.; Guillot, S.; Morsink, S. M.; Bilous, A. V.; Arzoumanian, Z.; Choudhury, D.; Deneva, J. S.; et al. A NICER View of the Massive Pulsar PSR J0740+6620 Informed by Radio Timing and XMM-Newton Spectroscopy. {\bf 2021}, arXiv: 2105.06980.
	
	\bibitem{nicer4} Miller, M. C.; Lamb, F. K.; Dittmann, A. J.; Bogdanov, S.; Arzoumanian, Z.; Gendreau, K. C.; Guillot, S.; Ho, W. C. G.; Lattimer, J. M.; Loewenstein, M.; et al. The Radius of PSR J0740+6620 from NICER and XMM-Newton Data. {\bf 2021}, arXiv: 2105.06979.
	
	\bibitem{ligo1} Abbott, B. P.; Abbott, R.; Abbott, T. D.; Acernese, F.; Ackley, K.; Adams, C.; Adams, T.; Addesso, P.; Adhikari, R. X.; Adya, V. B.; et al. GW170817: Observation of Gravitational Waves from a Binary Neutron Star Inspiral. {\em Phys. Rev. Lett.} {\bf 2017}, {\em 119}, 161101, DOI: 10.1103/PhysRevLett.119.161101.
	
	\bibitem{ligo2} Abbott, B. P.; Abbott, R.; Abbott, T. D.; Acernese, F.; Ackley, K.; Adams, C.; Adams, T.; Addesso, P.; Adhikari, R. X.; Adya, V. B.; et al. GW170817: Measurements of Neutron Star Radii and Equation of State. {\em Phys. Rev. Lett.} {\bf 2018}, {\em 121}, 161101, DOI: 10.1103/PhysRevLett.121.161101.
	
	\bibitem{ligo3} Abbott, B. P.; Abbott, R.; Abbott, T. D.; Acernese, F.; Ackley, K.; Adams, C.; Adams, T.; Addesso, P.; Adhikari, R. X.; Adya, V. B.; et al. Properties of the Binary Neutron Star Merger GW170817. {\em Phys. Rev. X} {\bf 2019}, {\em 9}, 011001, DOI: 10.1103/PhysRevX.9.011001.
	
	\bibitem{ligodec}  The LIGO Scientific Collaboration et al. Advanced LIGO. {\em Class. Quantum Grav.} {\bf 2015}, {\em 32}, 074001. DOI: 10.1088/0264-9381/32/7/074001.
	
	\bibitem{virgodec} Acernese, F.; Agathos, M.; Agatsuma, K.; Aisa, D.;
	Allemandou, N.;  Allocca, A.; Amarni, J.; Astone, P.; Balestri, G.; Ballardin, G.; et al. Advanced Virgo: a second-generation interferometric gravitational wave detector. {\em Class. Quantum Grav.} {\bf 2015}, {\em 32}, 024001. DOI: 10.1088/0264-9381/32/2/024001.
	
	{ 
		\bibitem{ligo4}Abbott, B. P.; Abbott, R.; Abbott, T. D.; Acernese, F.; Ackley, K.; Adams, C.; Adams, T.; Addesso, P.; Adhikari, R. X.; Adya, V. B.; et al. Prospects for observing and localizing gravitational-wave transients with Advanced LIGO, Advanced Virgo and KAGRA. {\em Living Rev Relativ} {\bf 2020}, {\em 23}, DOI:10.1007/s41114-020-00026-9.
	}
	
	\bibitem{Oertel2017} Oertel, M.; Hempel, M.; Klähn, T.; Typel, S. Equations of State for Supernovae and Compact Stars. {\em Rev. Mod. Phys.} {\bf 2017}, {\em 89}, 015007, DOI: 10.1103/RevModPhys.89.015007.
	
	\bibitem{Hartle1967} Hartle, J. B. Slowly Rotating Relativistic Stars. 1. Equations of Structure. {\em Astrophys. J.} {\bf 1967}, {\em 150}, 1005--1029. 
	
	\bibitem{Horowitz2019} Horowitz, C. J. Neutron Rich Matter in the Laboratory and in the Heavens after GW170817. {\em Annals of Physics}  {\bf 2019}, 411, 167992, DOI: 10.1016/j.aop.2019.167992.
	
	\bibitem{Burgio2020} Fiorella Burgio, G.; Vidaña, I. The Equation of State of Nuclear Matter: From Finite Nuclei to Neutron Stars. {\em Universe} {\bf 2020}, {\em 6}, 119, DOI: 10.3390/universe6080119.
	
	\bibitem{Guven2020} Güven, H.; Bozkurt, K.; Khan, E.; Margueron, J. Multimessenger and Multiphysics Bayesian Inference for the GW170817 Binary Neutron Star Merger. {\em Phys. Rev. C} {\bf 2020}, {\em 102}, 015805, DOI: 10.1103/PhysRevC.102.015805.
	
	\bibitem{prex} C. Gal, New precision measurement of the neutral weak
	form factor of 208Pb, Presented on behalf of the PREX Collaboration at the 2020 Meeting of the Division of Nuclear Physics of the American Physical Society.
	
	\bibitem{Piekarewicz2021}Reed, B. T.; Fattoyev, F. J.; Horowitz, C. J.; Piekarewicz, J. Implications of PREX-2 on the Equation of State of Neutron-Rich Matter. {\em Phys. Rev. Lett.} {\bf 2021}, {\em 126}, 172503, DOI: 10.1103/PhysRevLett.126.172503.
	
	\bibitem{Essick2021} Essick, R.; Tews, I.; Landry, P.; Schwenk, A. Astrophysical Constraints on the Symmetry Energy and the Neutron Skin of $^{208}$ Pb with Minimal Modeling Assumptions. {\bf 2021}, arXiv: 2102.10074.
	
	\bibitem{metamodel_2} Margueron, J.; Hoffmann Casali, R.; Gulminelli, F. Equation of State for Dense Nucleonic Matter from Metamodeling. II. Predictions for Neutron Star Properties.  {\em Phys. Rev. C} {\bf  2018}, {\em 97}, 025806, DOI: 10.1103/PhysRevC.97.025806.
	
	\bibitem{Steiner2013} Steiner, A. W.; Lattimer, J. M.; Brown, E. F. The Neutron Star Mass-Radius Relation and the Equation of State of Dense Matter. {\em Astrophys. J. Lett.} {\bf 2013}, {\em 765}, L5, DOI: 10.1088/2041-8205/765/1/L5.
	
	\bibitem{metamodel_1} Margueron, J.; Hoffmann Casali, R.; Gulminelli, F. Equation of State for Dense Nucleonic Matter from Metamodeling. I. Foundational Aspects. {\em Phys. Rev. C} {\bf 2018}, {\em 97}, 025805, DOI: 10.1103/PhysRevC.97.025805.
	
	\bibitem{Li2018} Zhang, N.-B.; Li, B.-A.; Xu, J. Combined Constraints on the Equation of State of Dense Neutron-Rich Matter from Terrestrial Nuclear Experiments and Observations of Neutron Stars. {\em Astrophys. J.} {\bf 2018}, {\em 859}, 90, DOI: 10.3847/1538-4357/aac027.
	
	\bibitem{Lim2019} Lim, Y.; Holt, J. W. Bayesian Modeling of the Nuclear Equation of State for Neutron Star Tidal Deformabilities and GW170817. {\em Eur. Phys. J. A} {\bf 2019}, {\em 55}, 209, DOI: 10.1140/epja/i2019-12917-9.
	
	\bibitem{Carreau2019}Carreau, T.; Gulminelli, F.; Margueron, J. General Predictions for the Neutron Star Crustal Moment of Inertia. {\em Phys. Rev. C} {\bf 2019}, {\em 100}, 055803,
	DOI: 10.1103/PhysRevC.100.055803.
	
	\bibitem{Tsang2020} Tsang, C. Y.; Tsang, M. B.; Danielewicz, P.; Lynch, W. G.; Fattoyev, F. J. Impact of the Neutron-Star Deformability on Equation of State Parameters. {\em Phys. Rev. C} {\bf 2020}, {\em 102}, 045808, DOI: 10.1103/PhysRevC.102.045808.
	
	\bibitem{CarreauEPJA} Carreau, T.; Gulminelli, F.; Margueron, J. Bayesian Analysis of the Crust-Core Transition with a Compressible Liquid-Drop Model. {\em Eur. Phys. J. A} {\bf 2019}, {\em 55}, 188, DOI: 10.1140/epja/i2019-12884-1.
	
	\bibitem{ame2016} Huang, W. J.; Audi, G.; Wang, M.; Kondev, F. G.; Naimi, S.; Xu, X. The AME2016 Atomic Mass Evaluation (I). Evaluation of Input Data; and Adjustment Procedures. {\em Chinese Phys. C}  {\bf 2017}, {\em 41}, 030002, DOI: 10.1088/1674-1137/41/3/030002.
	%\bibitem{ame2016}M.Wang, W. Huang, F. Kondev, G. Audi, and S. Naimi, Chin. Phys. C 45, 030003 (2021).
	\bibitem{Hoa2021} Dinh Thi, H.; Carreau, T.; Fantina, A. F.; Gulminelli, F. Uncertainties in the Pasta-Phase Properties in Catalysed Neutron Stars. {\em Astron. Astrophys.  2021, accepted},  DOI: https://doi.org/10.1051/0004-6361/202141192. 

 \bibitem{Ravenhall1983} Ravenhall, D. G.; Pethick, C. J.; Wilson, J. R. Structure of Matter below Nuclear Saturation Density. {\em Phys. Rev. Lett.} {\bf 1983}, {\em 50}, 2066, DOI: 10.1103/PhysRevLett.50.2066. 

\bibitem{Lattimer1991}  Lattimer, J. M.; Douglas Swesty, F. A Generalized Equation of State for Hot, Dense Matter. {\em Nucl. Phys. A}  {\bf 1991}, {\em 535}, 331--376, DOI: 10.1016/0375-9474(91)90452-C.

\bibitem{Lorenzet93} Lorenz, C. P.; Ravenhall, D. G.; Pethick, C. J. Neutron Star Crusts. {\em Phys. Rev. Lett.} {\bf 1993}, {\em 70}, 379--382, DOI: 10.1103/PhysRevLett.70.379.
	
\bibitem{Newton2013} Newton, W. G.; Gearheart, M.; Li, B. A. A Survey of the Parameter Space of the Compressible Liquid Drop Model as Applied to the Neutron Star Inner Crust. {\em The Astrophysical Journal Supplement Series} {\bf 2013}, {\em204}, 9, DOI: 10.1088/0067-0049/204/1/9.
	
\bibitem{Balliet21} Balliet, L. E.; Newton, W. G.; Cantu, S.; Budimir, S. Prior Probability Distributions of Neutron Star Crust Models. {\em Astrophys. J.} {\bf 2021}, {\em 918}, 79, DOI: https://doi.org/10.3847/1538-4357/ac06a4.

\bibitem{Fortin2016} Fortin, M.; Providência, C.; Raduta, A. R.; Gulminelli, F.; Zdunik, J. L.; Haensel, P.; Bejger, M. Neutron Star Radii and Crusts: Uncertainties and Unified Equations of State. {\em Phys. Rev. C} {\bf 2016}, {\em 94}, 035804, DOI: 10.1103/PhysRevC.94.035804.

\bibitem{Drischler2016} Drischler, C.; Hebeler, K.; Schwenk, A. Asymmetric Nuclear Matter Based on Chiral Two- and Three-Nucleon Interactions. {\em Phys. Rev. C} {\bf 2016}, {\em 93}, 054314, DOI: 10.1103/PhysRevC.93.054314.

\bibitem{Antoniadis2013} Antoniadis, J.; Freire, P. C. C.; Wex, N.; Tauris, T. M.; Lynch, R. S.; van Kerkwijk, M. H.; Kramer, M.; Bassa, C.; Dhillon, V. S.; Driebe, T.; et al. A Massive Pulsar in a Compact Relativistic Binary. {\em Science} {\bf 2013}, {\em 340}, DOI: 10.1126/science.1233232.

\bibitem{Fonseca2021} Fonseca, E.; Cromartie, H. T.; Pennucci, T. T.; Ray, P. S.; Kirichenko, A. Y.; Ransom, S. M.; Demorest, P. B.; Stairs, I. H.; Arzoumanian, Z.; Guillemot, L.; et al. Refined Mass and Geometric Measurements of the High-Mass PSR J0740+6620. {\em Astrophys. J. Lett.} {\bf 2021}, {\em 915}, L12, DOI: 10.3847/2041-8213/ac03b8.

\bibitem{Oppenheimer1939} Oppenheimer, J. R.; Volkoff, G. M. On Massive Neutron Cores. {\em Phys. Rev.} {\bf 1939}, {\em 55}, 374--381, DOI: 10.1103/PhysRev.55.374.

\bibitem{Tolman1939} Tolman, R. C. Static Solutions of Einstein’s Field Equations for Spheres of Fluid. {\em Phys. Rev.} {\bf 1939}, {\em 55}, 364--373, DOI: 10.1103/PhysRev.55.364.

\bibitem{dataligo} LIGO Document P1800370-v5: Parameter estimation sample release for GWTC-1, DOI: https://doi.org/10.7935/KSX7-QQ51.

\bibitem{Hinderer2008} Hinderer, T. Tidal Love Numbers of Neutron Stars. {\em Astrophys. J.} {\bf 2008}, {\em 677}, 1216--1220, DOI: 10.1086/533487.

\bibitem{Binnington2009} Binnington, T.; Poisson, E. Relativistic Theory of Tidal Love Numbers. {\em Phys. Rev. D}  {\bf 2009}, {\em 80}, 084018, DOI: 10.1103/PhysRevD.80.084018.

\bibitem{Damour2009} Damour, T.; Nagar, A. Relativistic Tidal Properties of Neutron Stars. {\em Phys. Rev. D} {\bf  2009}, {\em 80}, 084035, DOI: 10.1103/PhysRevD.80.084035.

\bibitem{Hoaepja2021} Dinh Thi, H.; Fantina, A. F.; Gulminelli, F. The Effect of the Energy Functional on the Pasta-Phase Properties of Catalysed Neutron Stars. {\em  Eur. Phys. J. A 2021, submitted}.

\bibitem{glitch} Espinoza, C. M.; Lyne, A. G.; Stappers, B. W.; Kramer, M. A Study of 315 Glitches in the Rotation of 102 Pulsars. {\em Mon. Not. R. Astron. Soc.}  {\bf 2011}, {\em 414}, 1679--1704, DOI: 10.1111/j.1365-2966.2011.18503.x.

\bibitem{Grill2012} Grill, F.; Providência, C.; Avancini, S. S. Neutron Star Inner Crust and Symmetry Energy. {\em Phys. Rev. C} {\em 2012}, {\em 85}, 055808, DOI:
10.1103/PhysRevC.85.055808.

\bibitem{Pearson2020} Pearson, J. M.; Chamel, N.; Potekhin, A. Y. Unified Equations of State for Cold Nonaccreting Neutron Stars with Brussels-Montreal Functionals. II. Pasta Phases in Semiclassical Approximation. {\em Phys. Rev. C} {\bf 2020}, {\em 101}, 15802, DOI: 
10.1103/PhysRevC.101.015802.

\bibitem{Lindblom2010} Lindblom, L. Spectral Representations of Neutron-Star Equations of State. {\em Phys. Rev. D} {\bf 2010}, {\em 82}, 103011, DOI: 10.1103/PhysRevD.82.103011.

\bibitem{SLy2001} Lattimer, J. M.; Prakash, M. Neutron Star Structure and the Equation of State. {\em Astrophys. J.} {\bf  2001}, {\em 550}, 426--442, DOI: 10.1086/319702.

{
	\bibitem{Landry2020}Landry, P.; Essick, R.; Chatziioannou, K. Nonparametric constraints on neutron star matter with existing and upcoming gravitational wave and pulsar observations. {\em Phys. Rev. D} {\bf 2020} {\em 101}, 123007, DOI: https://doi.org/10.1103/PhysRevD.101.123007.
}

\bibitem{Pang2021} Pang, P. T. H.; Tews, I.; Coughlin, M. W.; Bulla, M.; Broeck, C. Van Den; Dietrich, T. Nuclear-Physics Multi-Messenger Astrophysics Constraints on the Neutron-Star Equation of State: Adding NICER’s PSR J0740+6620 Measurement. {\bf 2021}, arXiv: 2105.08688.

\bibitem{Raaijmakers2021} Raaijmakers, G.; Greif, S. K.; Hebeler, K.; Hinderer, T.; Nissanke, S.; Schwenk, A.; Riley, T. E.; Watts, A. L.; Lattimer, J. M.; Ho, W. C. G. Constraints on the Dense Matter Equation of State and Neutron Star Properties from NICER’s Mass-Radius Estimate of PSR J0740+6620 and Multimessenger Observations. {\bf 2021}, arXiv: 2105.06981.

\bibitem{Goriely2013} Goriely, S.; Chamel, N.; Pearson, J. M. Further Explorations of Skyrme-Hartree-Fock-Bogoliubov Mass Formulas. XIII. the 2012 Atomic Mass Evaluation and the Symmetry Coefficient. {\em Phys. Rev. C} {\bf 2013}, {\em 88}, 024308, DOI: 10.1103/PhysRevC.88.024308.

\bibitem{Chabanat1998} Chabanat, E.; Bonche, P.; Haensel, P.; Meyer, J.; Schaeffer, R. A Skyrme Parametrization from Subnuclear to Neutron Star Densities Part II. Nuclei Far from Stabilities. {\em Nucl. Phys. A} {\bf 1998}, {\em 635}, 231–256. DOI: https://doi.org/10.1016/S0375-9474(98)00180-8.

\bibitem{Kuwabara1995} Sumiyoshi, K.; Kuwabara, H.; Toki, H. Relativistic mean-field theory with non-linear $\sigma$ and $\omega$ terms for neutron stars and supernovae. {\em Nucl. Phys. A} {\bf 1995}, {\em 581}, 725–746. DOI: 10.1016/0375-9474(94)00335-K.

\bibitem{Long2004} Long, W.; Meng, J.; Van Giai, N.; Zhou, S. G. New Effective Interactions in Relativistic Mean Field Theory with Nonlinear Terms and Density-Dependent Meson-Nucleon Coupling. {\em Phys. Rev. C } {\bf 2004}, {\em 69}, 034319. DOI: 10.1103/PhysRevC.69.034319.

\bibitem{Legred2021} Legred, I.; Chatziioannou, K.; Essick, R.; Han, S.; Landry, P. Impact of the PSR J0740+6620 Radius Constraint on the Properties of High-Density Matter. {\bf 2021}, arXiv: 2106.05313.

\bibitem{Adhikari2021} Adhikari, D.; Albataineh, H.; Androic, D.; Aniol, K.; Armstrong, D. S.; Averett, T.; Ayerbe Gayoso, C.; Barcus, S.; Bellini, V.; Beminiwattha, R. S.; et al. Accurate Determination of the Neutron Skin Thickness of $^{208}$Pb through Parity-Violation in Electron Scattering. {\em Phys. Rev. Lett.} {\bf 2021}, {\em 126}, 172502, DOI:10.1103/PhysRevLett.126.172502.

\bibitem{Biswas2021} Biswas, B. Impact of PREX-II and Combined Radio/NICER/XMM-Newton’s Mass-Radius Measurement of PSRJ0740+6620 on the Dense Matter Equation of State. {\bf 2021}, arXiv: 2105.02886.

\bibitem{Malik2018} Malik, T.; Alam, N.; Fortin, M.; Providência, C.; Agrawal, B. K.; Jha, T. K.; Kumar, B.; Patra, S. K. GW170817: Constraining the Nuclear Matter Equation of State from the Neutron Star Tidal Deformability. {\em Phys. Rev. C} {\bf 2018}, {\em 98}, 035804, DOI: 10.1103/PhysRevC.98.035804.

\bibitem{Fattoyev} Fattoyev, F. J.; Piekarewicz, J.; Horowitz, C. J. Neutron Skins and Neutron Stars in the Multimessenger Era. {\em Phys. Rev. Lett.} {\bf 2018}, {\em 120}, 172702, DOI: 10.1103/PhysRevLett.120.172702.

\bibitem{Annala2018} Annala, E.; Gorda, T.; Kurkela, A.; Vuorinen, A. Gravitational-Wave Constraints on the Neutron-Star-Matter Equation of State. {\em Phys. Rev. Lett.} {\bf 2018}, {\em 120}, 172703, DOI: 10.1103/PhysRevLett.120.172703.

\bibitem{Lourenco2019} Lourenço, O.; Dutra, M.; Lenzi, C. H.; Flores, C. V.; Menezes, D. P. Consistent Relativistic Mean-Field Models Constrained by GW170817. {\em  Phys. Rev. C} {\bf 2019}, {\em 99}, 045202, DOI: 10.1103/PhysRevC.99.045202.

\bibitem{Klahn2006} Klähn, T.; Blaschke, D.; Typel, S.; van Dalen, E. N. E.; Faessler, A.; Fuchs, C.; Gaitanos, T.; Grigorian, H.; Ho, A.; Kolomeitsev, E. E.; et al. Constraints on the High-Density Nuclear Equation of State from the Phenomenology of Compact Stars and Heavy-Ion Collisions. {\em Phys. Rev. C} {\bf 2006}, {\em 74}, 035802, DOI: 10.1103/PhysRevC.74.035802.
%\bibitem{CarreauThesis} T. Carreau, PhD dissertation, https://tel.archives-
%ouvertes.fr/tel-03019954 (2020).

\bibitem{Dobaczewski14} Dobaczewski, J.; Nazarewicz, W.; Reinhard, P.-G. Error estimates of theoretical models: a guide.
{\em Jour. Phys. G} {\bf 2014}, {\em 41}, 074001, DOI: 10.1088/0954-3899/41/7/074001.


\bibitem{Margueon-Gulminelli2019} Margueron, J.; Gulminelli, F. Effect of High-Order Empirical Parameters on the Nuclear Equation of State. {\em Phys. Rev. C} {\bf 2019}, {\em 99}, 025806, DOI: 10.1103/PhysRevC.99.025806.

\bibitem{Danielewicz2009} Danielewicz, P.; Lee, J. Symmetry Energy I: Semi-Infinite Matter. {\em  Nucl. Phys. A} {\bf  2009}, {\em 818}, 36--96,  DOI: 10.1016/j.nuclphysa.2008.11.007.

\bibitem{Chen2009} Chen, L.-W.; Cai, B.-J.; Ko, C. M.; Li, B.-A.; Shen, C.; Xu, J. Higher-Order Effects on the Incompressibility of Isospin Asymmetric Nuclear Matter. {\em  Phys. Rev. C} {\bf  2009}, {\em 80}, 014322, DOI: 10.1103/PhysRevC.80.014322.

\bibitem{Vidana2009} Vidaña, I.; Providência, C.; Polls, A.; Rios, A. Density Dependence of the Nuclear Symmetry Energy: A Microscopic Perspective. {\em Phys. Rev. C} {\bf 2009}, {\em 80}, 045806, DOI: 10.1103/PhysRevC.80.045806.

\bibitem{Ducoin2010} Ducoin, C.; Margueron, J.; Providência, C. Nuclear Symmetry Energy and Core-Crust Transition in Neutron Stars: A Critical Study. {\em  Europhysics Lett.} {\bf 2010}, {\em 91}, 32001, DOI: 10.1209/0295-5075/91/32001.

\bibitem{Santos2014}Santos, B. M.; Dutra, M.; Lourenço, O.; Delfino, A. Correlations between the Nuclear Matter Symmetry Energy, Its Slope, and Curvature from a Nonrelativistic Solvable Approach and Beyond. {\em Phys. Rev. C} {\bf  2014}, {\em  90}, 035203, DOI: 10.1103/PhysRevC.90.035203.



\bibitem{Mondal17}Mondal, C.; Agrawal, B. K.; De, J. N.; Samaddar, S. K.; Centelles, M.; Vi\~nas, X.
Interdependence of different symmetry energy elements. {\em Phys. Rev. C} {\bf  2017}, {\em  96}, 
021302, DOI: 10.1103/PhysRevC.96.021302.

\bibitem{Mondal18}Mondal, C.; Agrawal, B. K.; De, J. N.; Samaddar, S. K. Correlations among symmetry energy elements 
in Skyrme models. {\em Int. J. Mod. Phys. E} {\bf  2018}, {\em  27}, 
1850078, DOI: 10.1142/S0218301318500787.

\bibitem{HADES} HADES collaboration:  Adamczewski-Musch, J.; et al. {\em Phys. Rev. Lett.} {\bf 2017}, {\em 125}, 262301,  DOI: 10.1103/PhysRevLett.125.262301.


\end{thebibliography}
\end{document}